\documentclass[aps,prab,reprint,groupedaddress]{revtex4-2}

\usepackage{graphicx}

\usepackage{amsmath,amssymb}

\begin{document}

\title{Using Kernel-Based Statistical Distance to Study the Dynamics of Charged Particle Beams in Particle-Based Simulation Codes}

\author{Chad E. Mitchell}
\email{ChadMitchell@lbl.gov}
\author{Robert D. Ryne}
\address{Lawrence Berkeley National Laboratory, Berkeley, CA  94720, USA}

 \author{Kilean Hwang}
 \address{Facility for Rare Isotope Beams, Michigan State University, East Lansing, MI  48824, USA}

\date{\today}

\begin{abstract}
Measures of discrepancy between probability distributions (statistical distance) are widely used in the fields of artificial intelligence and machine learning.  We describe how certain measures of statistical distance can be implemented as numerical diagnostics for simulations involving charged-particle beams.  Related measures of statistical dependence are also described.  The resulting diagnostics provide sensitive measures of dynamical processes important for beams in nonlinear or high-intensity systems, which are otherwise difficult to characterize.  The focus is on kernel-based methods such as Maximum Mean Discrepancy, which have a well-developed mathematical foundation and reasonable computational complexity.  Several benchmark problems and examples involving intense beams are discussed.  While the focus is on charged-particle beams, these methods may also be applied to other many-body systems such as plasmas or gravitational systems.  
\end{abstract}

\pacs{}

\maketitle



\section{Introduction}
When modeling the dynamics of charged particle beams, the following question often arises.  Given two ensembles of simulation particles, how similar are they?
In particular, when do two particle ensembles represent the same underlying phase space density?  This question is central
to validating the random sampling algorithm used for initial beam generation, to comparing particle-based output across multiple simulation codes, to matching the particle beam successfully into a periodic transport system, and for studying the long-time phase space evolution of beams in circular and multi-pass systems.

Two beams are typically compared using their first and second moments, followed by qualitative examination of the beam phase space.  For beams at high intensity, in the presence of collective instabilities, or in the presence of highly nonlinear transport, the details of the distribution (including higher-order moments and nonlinear correlations) become increasingly important.  Such systems may exhibit filamentation of
the beam phase space, developing structure on finer and finer scales, in some cases relaxing to a quasi-stationary state.  It is then valuable to have quantitative particle-based diagnostics that can characterize the nonlinear dynamical processes of regular or chaotic mixing and collisionless relaxation \cite{Kandrup, Bohn1, UMER, LongRange, CoulombR, Levin1}.

These problems can best be addressed by implementing a two-sample measure of statistical distance with well-understood mathematical properties, which can be used to compare particle populations.  Statistical distances are widely used in machine learning (ML), information theory, statistics, probability theory, and data mining.  Unfortunately, such quantities tend to have high computational complexity.  For beam physics applications, the complexity must scale well with the number of particles $n$ (as $O(n^2)$ or better) and with the phase space dimension $d$ (up to at least $d=6$).

We describe the kernel-based statistical distance known as Maximum Mean Discrepancy (MMD) \cite{MMDdef}, which has recently had a major impact in the ML community.  Broadly speaking, kernel methods allow nonlinear problems involving higher-order statistics to be treated using linear methods, by embedding the set of probability distributions into a reproducing kernel Hilbert space (RKHS) \cite{RKHStext, RKHSsurvey}.  We do not discuss the RKHS formalism, although many results are most naturally viewed in this context.
The use of MMD as a statistical distance leads naturally to a measure of statistical dependence or correlation, known here as the Hilbert Schmidt Correlation (HSCor) \cite{HSIC}.  These two diagnostics can provide powerful quantitative tools to study the beam dynamics processes mentioned above.

The structure of this paper is as follows.  In Section II we review several concepts of statistical distance.  Section III describes the properties of Maximum Mean Discrepancy and its implementation in particle-based tracking codes.  Section IV describes the properties of the Hilbert Schmidt Correlation and its corresponding implementation.  In Section V, we demonstrate how these tools can be used for beam dynamics applications, using several benchmark problems involving symplectic maps.  In Section VI, we apply these tools to examples involving high-intensity beams with self-consistent space charge.  Section VII contains a summary and conclusions.  There are two Appendices.

\section{Statistical Distance}\label{sec:distance}
Consider a beam described by its kinetic distribution function $f$ on an appropriate phase space of dimension $d$ with coordinates $X=(q_1,p_1,\ldots,q_{d/2},p_{d/2})$, normalized so that
\begin{equation}
\int f(X)dX=1. \label{normalization}
\end{equation}
Then $f$ defines a probability density on the phase space.  By Liouville's theorem, the condition (\ref{normalization}) is preserved during the beam evolution, and we often use $f_t$ to denote the phase space density at time $t$.

We wish to define a distance $\rho$ between pairs of probability densities $f$ and $g$, such that $\rho$ satisfies the following natural conditions:
\begin{subequations}\label{metric}
\begin{align*}
&{\rm i}) \text{ non-negativity:  } \rho(f,g)\geq 0, \\
&{\rm ii}) \text{ symmetry:  } \rho(f,g)=\rho(g,f), \\
&{\rm iii}) \text{ the triangle inequality:  } \rho(f,g)\leq \rho(f,h)+\rho(h,g) \notag \\
&\text{   for any probability density $h$}, \\
&{\rm iv}) \text{ identity:  }\rho(f,g)=0\text{ if and only if $f=g$ } \notag \\ 
&\text{(except on a possible set of zero probability)}. \label{criterion4}
\end{align*}
\end{subequations}
That is, $\rho$ should define a metric on the set of all probability densities \cite{DensityDef}.

The distance should also capture the concept of relaxation of beams, so that $f_t$ relaxes to $f_{eq}$ in the ``coarse-grained" sense if and only if $\rho(f_t,f_{eq})\rightarrow 0$ as $t\rightarrow\infty$.
The concept of relaxation in the ``coarse-grained" sense is well-formalized by the probabilistic concept of weak convergence \cite{Dudley, Billingsley}:  we say $f_t\Rightarrow f_{eq}$ if and only if:
\begin{equation}
\lim_{t\rightarrow\infty}\int f_t(X)\phi(X)dX = \int f_{eq}(X)\phi(X)dX \label{coarsegrained}
\end{equation}
for each bounded, continuous function $\phi$ on the phase space.  Informally, (\ref{coarsegrained}) states that the ensemble average of each well-behaved observable $\phi$ must approach the ensemble average of $\phi$ over $f_{eq}$ as $t\rightarrow\infty$.  This is a natural concept of convergence to use when characterizing the long-time behavior of beams.

The description of the beam as a dynamically evolving probability density is valid provided that the beam consists of a single particle species, without charge loss.  In the presence of charge loss, one is often interested in the behavior of the part of the beam that lies within a bounded subregion of the phase space.  The tools described in this paper can also be adapted to treat this case, using a technique to be described in Section \ref{sec:dynamic}.

In the following subsections, we briefly describe several indicators of statistical distance that are used in ML and pattern recognition, information science, probability, and statistics.  For a general survey, see for example \cite{Statdist}.

\subsection{Kullback-Liebler Divergence}
The most widely-used statistical distance is the Kullback-Liebler (KL) divergence $D_{KL}$, given by:
\begin{equation}
D_{KL}(f||g)=\int f(X)\log\frac{f(X)}{g(X)}dX, \label{KLeq}
\end{equation}
which originated in information theory as a measure of the relative entropy of one probability density with respect to another \cite{KLdef, KLbook}.
It has the property that $D_{KL}(f||g)\geq 0$, with equality if and only if $f=g$.  (It satisfies metric conditions i and iv above.)  However, the integral in (\ref{KLeq}) is not defined for all pairs of probability densities $f$ and $g$, and $D_{KL}$ fails to satisfy conditions ii and iii, so it is not a metric in the above sense. 

Note that $D_{KL}$ is invariant under any symplectic time-evolution map $\mathcal{M}_t$, since:
\begin{align}
D_{KL}(f_t||g_t)&=\int f_t(X)\log\frac{f_t(X)}{g_t(X)}dX  \\
&=\int f_0(\mathcal{M}_t^{-1}(X))\log\frac{f_0(\mathcal{M}_t^{-1}(X))}{g_0(\mathcal{M}_t^{-1}(X))}dX  \notag \\
&=\int f_0(X')\log\frac{f_0(X')}{g_0(X')}dX'=D_{KL}(f_0||g_0), \notag
\end{align}
where we used the fact that $\mathcal{M}_t$ has Jacobian determinant 1.  In particular, if a density $f_{eq}$ is invariant under $\mathcal{M}_t$, in the sense that
\begin{equation}
f_{eq}(\mathcal{M}_t^{-1}(X))=f_{eq}(X),
\end{equation}
then $D_{KL}(f_t||f_{eq})$ is independent of $t$.  Thus, $D_{KL}$ cannot capture weak convergence of the form $f_t\Rightarrow f_{eq}$ to an invariant density $f_{eq}$ under a symplectic time-evolution.

The KL divergence has been used in kinetic simulations \cite{KLapp}, and it is well-motivated by statistical mechanics considerations.  However, typical algorithms for computing $D_{KL}$ require binning the (possibly highly-filamented) distribution functions $f$ and $g$, which becomes increasingly problematic in a phase space of dimension $>2$.
Although gridless two-sample estimation algorithms also exist \cite{KLest}, the rate of convergence with particle number can be arbitrarily slow, with a convergence rate that varies with the distribution.  This makes the quantity $D_{KL}$ difficult to estimate reliably from samples, especially in spaces of high dimension.

\subsection{Wasserstein Metric}
The $p$-Wasserstein distance $(p=1,2,\ldots)$ is defined by:
\begin{equation}
W_p(f,g)=\left(\min_{h}\int |X-Y|^p h(X,Y)dX dY\right)^{1/p} \label{Wassdef}
\end{equation}
where $|X-Y|$ denotes the Euclidean distance between points $X$ and $Y$.
Here the minimum is taken \cite{note1} over all joint probability densities $h$ with marginal densities $f$ and $g$, so that:
\begin{equation}
f(X)=\int h(X,Y)dY,\quad g(Y)=\int h(X,Y)dX.
\end{equation}
The distances $W_p$ originated in the theory of optimal transport \cite{OptTransp, Wassdef1, Wassdef2}, where
the case $p=1$ is also known as the Kantorovich-Rubinstein metric or the Earth Mover's Distance (EMD).  

Note that (\ref{Wassdef}) is guaranteed to be finite when the densities $f$ and $g$ both have finite moments of order $p$.  On the
set of all such densities, $W_p$ is known to satisfy all the metric conditions (i-iv).
It is also known that $W_p$ correctly captures the concept of weak convergence, in the sense that $W_p(f_t,f_{eq})\rightarrow 0$ (as $t\rightarrow\infty$) if and only if $f_t\Rightarrow f_{eq}$ and the $p$th moments of $f_t$ converge to those of $f_{eq}$.

Due to its desirable geometric properties, the Wasserstein distance has been applied in ML to tasks such as shape matching, image retrieval, graphics, and to the statistical analysis of detector events in high-energy colliders \cite{WassHEP, WassHEP2, WassHEP3}.  However, the estimation of $W_p$ from sample data requires solving a linear optimization problem with a computational complexity of $O(n^3\log n)$, where $n$ is the number of samples \cite{MMD3, MMD4}.  Furthermore, the sample estimate converges to the population value as $O(n^{-1/d})$ for $d>2$.  This makes $W_p$ challenging to use for practical beam dynamics simulations, which typically require $n\geq 10^5$ and $d\geq 4$.  Since there are few cases in which (\ref{Wassdef}) can be determined in closed form, algorithms for computing $W_p$ are also difficult to benchmark.

\subsection{Maximum Mean Discrepancy}
A {\it kernel} $k$ is a symmetric, real-valued function defined on pairs of phase space points that is positive definite, in the sense that:
\begin{equation}
\sum_{i=1}^N\sum_{j=1}^N c_ic_jk(X_i,X_j)\geq 0, \label{posdef}
\end{equation}
for each $N=1,2,\ldots$, each finite set of points $X_1,\ldots,X_N$ and real numbers $c_1,\ldots,c_N$.  Typical examples are provided in Appendix A.

To each kernel $k$ is associated a Hilbert space (RKHS) consisting of real-valued functions on the phase space.   The Maximum Mean Discrepancy (MMD) between two probability densities $f$ and $g$ is then defined by \cite{MMDdef}:
\begin{equation}
\gamma_k(f,g)=\max_{\phi}\left|\int f(X)\phi(X)dX-\int g(X)\phi(X)dX\right| \label{MMDdef}
\end{equation}
where the maximum is taken over all functions $\phi$ in the RKHS with $||\phi||_k\leq 1$, $||\cdot ||_k$ denotes the Hilbert space norm, and $|\cdot |$ denotes the absolute value.

When $k$ is bounded, the quantity in (\ref{MMDdef}) is defined for all probability densities $f$ and $g$, and $\gamma_k$ satisfies the metric conditions (i-iii).
Additional restrictions on $k$ are used to ensure that $\gamma_k$ satisfies condition (iv), and that $\gamma_k$ captures weak convergence, in the sense previously described.
 The class of kernels satisfying these restrictions has been extensively studied \cite{MMD1, MMD1b, MMD2, MMD2b, MMD2c}, and it includes most of the kernels widely used in ML, including those described in Appendix A.  
  
 Due to its well-developed mathematical foundation, its connection to other ML kernel methods such as support vector machines \cite{RKHSsurvey}, its applicability to domains more general than Euclidean space, and its relative ease of estimation, the distance $\gamma_k$ has become a powerful tool in statistical two-sample (homogeneity) testing for ML applications.  Estimation of (\ref{MMDdef}) from sample data can be achieved using $O(n^2)$ operations, and the sample estimate converges to the population value as $O(n^{-1/2})$, independently of the dimension $d$.  Furthermore, approximations also exist that can be computed with complexity $O(n)$, making $\gamma_k$ a practical quantity for beam dynamics applications.

\section{Properties of Maximum Mean Discrepancy}\label{sec:mmd}
Given a kernel $k$, the maximum appearing in (\ref{MMDdef}) can be evaluated exactly by using the properties of its corresponding RKHS.  As a result, the MMD between two probability densities $f$ and $g$ can be expressed using the explicit integral formula \cite{MMD1}:
\begin{equation}
\gamma_k(f,g)=\left(\iint k(X,X')\Delta(X)\Delta(X')dXdX'\right)^{1/2}, \label{MMDint}
\end{equation}
where $\Delta=f-g$.  
For certain choices of $k$, the quantity (\ref{MMDint}) coincides with other well-known indicators of statistical distance.
For example, in the special case that $k(X,X')=|X|+|X'|-|X-X'|$, (\ref{MMDint}) appears in the statistics literature as the {\it energy distance} \cite{Edist, EdistMMD}.

\subsection{Basic properties}
Given any kernel $k$, one may construct a corresponding kernel $k_N$ by:
\begin{equation}
k_N(X,X')=\frac{k(X,X')}{\sqrt{k(X,X)k(X',X')}}. \label{knormal}
\end{equation}
The condition that $k$ be positive definite (\ref{posdef}) then implies that $k_N$ is positive definite with $|k_N|\leq 1$.  For simplicity, we will assume that all kernels are so normalized.  
It then follows from (\ref{MMDint}) that the distance
$\gamma_k$ is dimensionless with:
\begin{equation}
0\leq \gamma_k\leq 2.
\end{equation}
It is natural to choose a kernel that reflects the underlying properties of the domain, so
we often consider kernels that are translation-invariant, in the sense that $k(X,X')=k(X+\delta X,X'+\delta X)$ for
any phase space displacement $\delta X$.
A continuous, translation-invariant kernel can be written in terms of its Fourier components as:
\begin{equation}
k(X,X')=\int e^{i(X-X')\cdot\omega}\Lambda(\omega)d\omega. \label{ktranslation}
\end{equation}
When $k$ is normalized, $\Lambda$ is a probability density \cite{DensityDef} on the space of frequencies $\omega=(\omega_1,\ldots,\omega_d)$.  Using (\ref{ktranslation}) in (\ref{MMDint}), one finds that:
\begin{equation}
\gamma_k(f,g)=\left(\int\left|\phi_f(\omega)-\phi_g(\omega)\right|^2\Lambda(\omega)d\omega\right)^{1/2}, \label{MMDft}
\end{equation}
where $\phi_f$ and $\phi_g$ denote the Fourier transforms of the densities $f$ and $g$, normalized so that:
\begin{equation}
\phi_f(\omega)=\int e^{iX\cdot\omega}f(X)dX.
\end{equation}
When the probability density $\Lambda$ is also an integrable function that is strictly positive everywhere, it is possible to prove that (\ref{MMDft}) satisfies the metric conditions (i-iv) and
correctly reflects the weak convergence of probability distributions, as previously described.

One additional property of $\gamma_k$ is also useful.
If $\{e_l:l=1,2,\ldots\}$ denotes an orthonormal basis for the RKHS associated with the kernel $k$, then we may define a complete sequence of beam ``moments" $m_l$ by:
\begin{equation}
m_l(f)=\int e_l(X)f(X)dX\quad (l=1,2,\ldots). \label{moments}
\end{equation}
It follows from (\ref{MMDint}) that the MMD between two distributions $f$ and $g$ may be written in terms of these moments as:
\begin{equation}
\gamma_k(f,g)=\sqrt{\sum_{l=1}^{\infty}\left|m_{l}(f)-m_{l}(g)\right|^2}. \label{MMDmoment}
\end{equation}
In particular, for every $l=1,2,\ldots$ we have:
\begin{equation}
\left|m_{l}(f)-m_{l}(g)\right|\leq \gamma_k(f,g).
\end{equation}
Thus when $\gamma_k$ is small, all of the moments $m_l$ of the distributions $f$ and $g$ must nearly coincide.  (An example is provided in Appendix A.)

\subsection{Sample estimate}
A direct estimate of (\ref{MMDint}) from particle (sample) data is given by \cite{MMDdef}:
\begin{align}
\gamma_k^2(f,g)&=\frac{1}{m^2}\sum_{i,j=1}^mk(X_i,X_j)-\frac{2}{mn}\sum_{i,j=1}^{m,n}k(X_i,Y_j) \notag \\
&+\frac{1}{n^2}\sum_{i,j=1}^nk(Y_i,Y_j), \label{MMDsample}
\end{align}
where the $m$ particle phase space coordinates $\{X_j\}_{j=1}^m$ are sampled from the distribution $f$, and the $n$ particle phase space coordinates $\{Y_j\}_{j=1}^n$ are
sampled from the distribution $g$.  Note that we allow $m\neq n$.

An alternative grouping of the sum (\ref{MMDsample}) that yields superior numerical performance in practice is given by:
\begin{equation}
\gamma_k^2(f,g)=\sum_{i,j=1}^{m+n}c_ic_jk(\hat{X}_i,\hat{X}_j), \label{alt}
\end{equation}
where 
\begin{equation}
c_j=\begin{cases}
\frac{1}{m}, & 1\leq j\leq m \\
-\frac{1}{n}, & m+1\leq j\leq m+n
\end{cases}\label{cjdef}
\end{equation}
and $\{\hat{X}_j\}_{j=1}^{m+n}$ contains the phase space coordinates sampled from the distribution $f$, followed by the phase space coordinates sampled from $g$, so that:
\begin{equation}
\hat{X}_j=\begin{cases}
X_j, & 1\leq j\leq m \\
Y_{j-m}, & m+1\leq j\leq m+n
\end{cases}.\label{Xjdef}
\end{equation}
In the form (\ref{alt}), it is clear from (\ref{posdef}) that the estimate satisfies $\gamma_k^2\geq 0$, and that the computational complexity is $O((m+n)^2)$.

In the special case that the kernel $k$ is translation-invariant, the complexity can be reduced by using the spectral representation (\ref{MMDft}) to approximate $\gamma_k$ as the sum \cite{RFF2, RFF3}:
\begin{equation}
\gamma_k^2(f,g)=\frac{1}{L}\sum_{l=1}^L\left| \sum_{j=1}^{m+n}c_je^{i\omega_l\cdot \hat{X}_j}\right|^2 \label{MMDfast}
\end{equation}
where the $\{\omega_l\}_{l=1}^L$ denote $L$ frequency vectors that are randomly sampled from the spectral probablity density $\Lambda$, and the quantities $\{c_j\}_{j=1}^{m+n}$ and $\{\hat{X}_j\}_{j=1}^{m+n}$ are given by (\ref{cjdef}-\ref{Xjdef}).
The complexity of (\ref{MMDfast}) is $O(L(m+n))$, where in practice it is sufficient to use $L\ll m+n$.  This results in significant speed-up when translation-invariant kernels are used to compute $\gamma_k$.

\subsection{Statistical error}
It is shown in \cite{MMD3, MMD4} that the sample estimate (\ref{MMDsample}) converges to the population value (\ref{MMDint}) as $O(m^{-1/2}+n^{-1/2})$ when $m,n\rightarrow\infty$, independently of the dimension of the underlying space.  A similar result \cite{RFF2, RFF3} may be obtained for the estimate (\ref{MMDfast}). 

In practice, the sample estimate in (\ref{MMDsample}-\ref{alt}) or (\ref{MMDfast}) is used to test the hypothesis that two distributions are distinct, $f\neq g$.  To set a confidence threshold for this test, one must know the statistical distribution of (\ref{MMDsample}) 
under the null hypothesis that $f=g$.  This problem has been treated in detail \cite{MMDdef, EdistMMD}.  For our purposes, it is enough to know that the rms value of $\gamma_k$ under the hypothesis that $f=g$ is given by taking the expected value of (\ref{MMDsample}), which is given by:
\begin{equation}
\gamma_k^{\rm noise}=\operatorname{E}[\gamma_k^2]^{1/2}=\sqrt{\frac{m+n}{mn}}\left(1-||f||_k^2\right)^{1/2}, \label{MMDnoise}
\end{equation}
where the notation $||\cdot||_k$ denotes
\begin{equation}
||f||_k^2=\iint k(X,X')f(X)f(X')dXdX'. \label{normk}
\end{equation}
An identical result is obtained by using (\ref{MMDfast}).
Thus (\ref{MMDnoise}) represents a statistical noise level for $\gamma_k$ that is associated with the use of a finite number of particles, and when $m=n$, we see that $\gamma_k^{\rm noise}\propto 1/\sqrt{n}$.
The probability $P$ that $\gamma_k$ exceeds a threshold value $\tau>0$ is then bounded above by Markov's inequality, which states that for any $\tau>0$:
\begin{equation}
P(\gamma_k>\tau)\leq \left(\frac{\gamma_k^{\rm noise}}{\tau}\right)^2. \label{Markov}
\end{equation}
It follows that (given the null hypothesis) large deviations above the noise floor $\gamma_k^{\rm noise}$ are rare.  A more detailed investigation \cite{MMDdef} reveals that the bound in (\ref{Markov}) is loose--that is, the probability on the left-hand side in (\ref{Markov}) is much smaller and decays more quickly with $\tau$ than (\ref{Markov}) alone would suggest.

\subsection{Numerical implementation}
The expressions in (\ref{alt}) and (\ref{MMDfast}) are straightforward to implement in a parallel particle-based beam dynamics simulation code.  The resulting numerical diagnostic, which we denote by MMD, may be used to compare an evolving particle distribution $f_t$ with itself after successive $t$-intervals $\Delta t$ ({\it e.g.}, lattice periods), or to compare the evolving particle distribution against a fixed reference distribution (usually an initial or predicted final distribution).  The frequency samples $\{\omega_l\}_{l=1}^L$ in (\ref{MMDfast}) may be generated once and stored at initialization of the simulation, or the samples may be drawn independently at each evaluation of the MMD.  The latter is the approach favored in the literature.

An algorithm can also be implemented to estimate $\gamma_k$ using the representation given in (\ref{MMDmoment}).  When the basis functions $\{e_l\}_{l=1}^{\infty}$ are known, it is straightforward to estimate the moments (\ref{moments}) from particle samples.  However, the number of basis functions required to obtain convergence of the sum (\ref{MMDmoment}) grows rapidly with the phase space dimension $d$.  For a Gaussian kernel with $d\geq 2$, we found that this algorithm was outperformed by the spectral algorithm (\ref{MMDfast}) for all examples tested.

A remark about the choice of kernel $k$ is in order.  For simplicity, we use a Gaussian kernel (\ref{kGaussian}) along each unbounded phase space coordinate.  The kernel width $\sigma$ is chosen to coincide with a typical rms beam size.  In some cases, it is natural to consider periodic domains (for example, if one models a longitudinal beam slice with periodic boundary conditions).  Along a phase space coordinate that is naturally periodic, we use a Poisson kernel (\ref{kPoisson}) with the appropriate periodicity.  For dynamical problems, it is important that the kernel remain fixed throughout the simulation.

\section{Hilbert Schmidt Correlation}\label{sec:hscor}
Recall that two random variables $X$ and $Y$ described by a joint probability density $P_{XY}$ are said to be independent when $P_{XY}=P_XP_Y$, where $P_X$ and $P_Y$ are the marginal densities given by:
\begin{subequations}
\begin{align}
P_X(X)&=\int P_{XY}(X,Y)dY, \\
P_Y(Y)&=\int P_{XY}(X,Y)dX.
\end{align}
\end{subequations}
Given a metric $\rho$ on the set of such joint probability densities, a natural measure of deviation from independence between $X$ and $Y$ is given by the distance $\rho(P_{XY},P_XP_Y)$.
This motivates the following definitions.

\subsection{Definition and properties}
For simplicity, we assume that $X$ and $Y$ take their values in the same space $\mathbb{R}^d$, on which a kernel $k$ is defined.  We define a new kernel $\kappa$ on $\mathbb{R}^d\times \mathbb{R}^d$ by:
\begin{equation}
\kappa((X,Y),(X',Y'))=k(X,X')k(Y,Y').
\end{equation}
It follows that $\kappa$ is symmetric and positive definite (\ref{posdef}).

The Hilbert Schmidt correlation $\mathcal{R}_k$ between $X$ and $Y$ is then defined by:
\begin{equation}
\mathcal{R}_k^2(X,Y)=\frac{\gamma_{\kappa}^2(P_{XY},P_XP_Y)}{\gamma_{\kappa}(P_{XX},P_XP_X)\gamma_{\kappa}(P_{YY},P_YP_Y)}. \label{HCor}
\end{equation}
The quantity in the numerator is known as the Hilbert Schmidt Independence Criterion (HSIC) \cite{HSIC, HSICb, HSIC2}.  The normalizing factor in the denominator appears in \cite{EdistMMD}, and is designed to ensure that $\mathcal{R}_k\leq 1$.  The joint densities $P_{XX}$ and $P_{YY}$ represent the limiting case of perfect correlation when $X=Y$, namely:
\begin{align}
P_{XX}(X,Y)&=P_X(X)\delta(Y-X), \label{PXXeq} \\
P_{YY}(X,Y)&=P_Y(Y)\delta(X-Y). \notag
\end{align}

It may be shown that the quantity in (\ref{HCor}) satisfies:
\begin{align}
&0\leq \mathcal{R}_k\leq 1, \label{Rkproperties} \\
&\mathcal{R}_k=0\text{ if and only if }X\text{ and }Y\text{ are independent}, \notag \\
&\mathcal{R}_k=1\text{ if }X\text{ and }Y\text{ are identical}. \notag
\end{align}
It follows that $\mathcal{R}_k$ provides a natural measure of (possibly nonlinear) correlation between $X$ and $Y$.
The special case when $k(X,X')=|X|+|X'|-|X-X'|$ is known in the statistics literature as the {\it distance correlation} (dCor) \cite{EdistMMD, dCor, Edist, dCorMetric}.

When the kernel $k$ is translation-invariant, we may use the representation (\ref{MMDft}) to write the numerator of (\ref{HCor}) as:
\begin{align}
&\gamma_{\kappa}^2(P_{XY},P_XP_Y)=  \notag \\
&\int\left|\phi_{XY}(\omega,\omega')-\phi_X(\omega)\phi_Y(\omega')\right|^2\Lambda(\omega)\Lambda(\omega')d\omega d\omega', \label{HSICft}
\end{align}
where $\Lambda$ is the spectral density of $k$ defined in (\ref{ktranslation}), and
\begin{subequations}
\begin{align}
\phi_{XY}(\omega,\omega')=\int e^{i\left(\omega\cdot X+\omega'\cdot Y\right)}P_{XY}(X,Y)dXdY, \\
\phi_{X}(\omega)=\phi_{XY}(\omega,0),\quad\quad \phi_Y(\omega')=\phi_{XY}(0,\omega').
\end{align}
\end{subequations}
Corresponding expressions for the factors in the denominator of (\ref{HCor}) are obtained from (\ref{HSICft}) by taking $Y\mapsto X$ or $X\mapsto Y$ as appropriate, and using (\ref{PXXeq}) to write:
\begin{subequations}
\begin{align}
\phi_{XX}(\omega,\omega')&=\int e^{i\left(\omega\cdot X+\omega'\cdot Y\right)}P_{XX}(X,Y)dXdY  \notag \\
&=\phi_X(\omega+\omega'), \\
\phi_{YY}(\omega,\omega')&=\int e^{i\left(\omega\cdot X+\omega'\cdot Y\right)}P_{YY}(X,Y)dXdY \notag \\
&=\phi_Y(\omega+\omega').
\end{align}
\end{subequations}

\subsection{Sample estimate}
An estimate of the numerator of (\ref{HCor}) from particle (sample) data is given by:
\begin{align}
\gamma_{\kappa}^2(P_{XY},P_XP_Y)&=\frac{1}{m^2}\sum_{i,j=1}^mk(X_i,X_j)k(Y_i,Y_j) \notag \\
&+\frac{1}{m^4}\sum_{i,j,q,r=1}^mk(X_i,X_j)k(Y_q,Y_r)  \notag \\
&-\frac{2}{m^3}\sum_{i,j,q=1}^m k(X_i,X_j)k(Y_i,Y_q) \label{HSICquad}
\end{align}
where the $m$ pairs $\{(X_j,Y_j)\}_{j=1}^m$ are sampled from $P_{XY}$.  The corresponding expressions appearing in the denominator of (\ref{HCor}) are obtained from (\ref{HSICquad}) by
taking $Y_{j}\mapsto X_{j}$ and $X_{j}\mapsto Y_{j}$, respectively.
It can be shown that (\ref{HSICquad}) can be computed with $O(m^2)$ complexity by introducing $O(m)$ storage.

In the special case that the kernel $k$ is translation invariant, this complexity can be further reduced by working in frequency space using (\ref{HSICft}). An efficient estimate is given by:
\begin{align}
\gamma_{\kappa}^2&(P_{XY}, P_XP_Y)=\frac{1}{L}\sum_{k=1}^L\left|\frac{1}{m}\sum_{j=1}^me^{i\omega_{2k-1}\cdot X_j}e^{-i\omega_{2k}\cdot Y_j} \right. \label{RFF} \\
&\left.-\left(\frac{1}{m}\sum_{j=1}^me^{i\omega_{2k-1}\cdot X_j}\right)\left(\frac{1}{m}\sum_{j=1}^me^{-i\omega_{2k}\cdot Y_j}\right)\right|^2, \notag
\end{align}
where $\{\omega_k\}_{k=1}^{2L}$ denote $2L$ frequency vectors that are randomly sampled from the spectral probability density $\Lambda$ associated with $k$.  
Note that the complexity of (\ref{RFF}) is given by $O(mL)$.   (This is a variant of the random Fourier features estimate appearing in \cite{RFF, RFF3}.)

\subsection{Statistical error}
In practice, the sample estimate in (\ref{HSICquad}) or (\ref{RFF}) is used to test the hypothesis that two random variables $X$ and $Y$ are independent.  To set a confidence threshold for this test, one needs to know the statistical distribution of (\ref{HSICquad}) under the null hypothesis that $P_{XY}=P_XP_Y$.  This problem has been treated in detail \cite{HSIC, HSICb, HSIC2, EdistMMD}.  For our purposes, it is enough to know the rms value of $\mathcal{R}_k$ under the hypothesis that $X$ and $Y$ are independent.  This is given by taking the expected value of (\ref{HSICquad}), which yields:
\begin{align}
&\operatorname{E}[\gamma_k^2(P_{XY},P_XP_Y)] \notag \\
&\quad\quad=\frac{(m-1)}{m^2}\left(1-||P_X||_k^2\right)\left(1-||P_Y||_k^2\right),
\end{align}
where $||\cdot||_k$ has the same meaning as in (\ref{normk}).
Thus we have, to leading order in $1/m$:
\begin{equation}
\mathcal{R}_k^{\rm noise}=\frac{1}{\sqrt{m}}\frac{\left(1-||P_X||_k^2\right)^{1/2}\left(1-||P_Y||_k^2\right)^{1/2}}{\gamma_{\kappa}(P_{XX},P_XP_X)\gamma_{\kappa}(P_{YY},P_YP_Y)}.
\label{HSICnoise}
\end{equation}
Note that (\ref{HSICnoise}) is fully determined by the marginal densities $P_X$ and $P_Y$ through (\ref{PXXeq}).  

Given the null hypothesis, an inequality corresponding to (\ref{Markov}) holds after $\gamma_k$ is replaced by $\mathcal{R}_k$, indicating that large deviations above the noise floor value $\mathcal{R}_k^{\rm noise}$ are rare.

\subsection{Numerical Implementation}
In practice, the random variables $X$ and $Y$ described above may represent two phase space coordinates within a single beam ({\it e.g.}, $X=z$ and $Y=p_z$) or two $d$-tuples of phase space coordinates ({\it e.g.}, $X=(x,p_x)$ and $Y=(y,p_y)$).  In this case, computation of $\mathcal{R}_k(X,Y)$ using (\ref{HSICquad}) or (\ref{RFF}) returns a measure of correlation between phase space coordinates within the beam.

Alternatively, let $X$ denote the vector of phase space coordinates for a particle within the beam at initial time (or lattice location) $t=0$, and let $Y$ denote the vector of phase space coordinates for the same particle at a later time $t$.  Then $\mathcal{R}_k(X,Y)$ measures the correlation of a particle's coordinates at time $t$ with the particle's initial coordinates, and this quantity will be denoted $\mathcal{R}_k(t)$ for simplicity.  The dynamical evolution of this quantity is intimately related to mixing.  (See Appendix B.)

Numerical evaluation of $\mathcal{R}_k(t)$ requires that each particle be assigned a unique index $j$, so that one may construct the particle pairs $\{(X_j,Y_j)\}_{j=1}^m$ at each desired evaluation time $t$. In particular, the arrays containing the initial and final phase space coordinates of particle $j$ must be stored on the same processor, which requires appropriate bookkeeping and possible communication.

\section{Applications to Idealized and Exactly-Solvable Models}

In this section, we illustrate several applications of the above tools to dynamical problems involving beams:  1) to compare two beams for benchmarking and quantifying beam mismatch, 2) to detect nonlinear phase space correlations and coupling between phase planes, 3) to verify numerically that a beam that is generated from a well-matched distribution remains stationary, 4) to study the relaxation of a non-equilibrium beam to a stationary state, and 5) to measure the rate of chaotic mixing and decay of correlations within a beam during its evolution.

To aid in benchmarking the diagnostics MMD $(\gamma_k$) and HSCor $(\mathcal{R}_k)$, idealized distributions and exactly-solvable models involving maps are used.  The next section will discuss realistic examples involving high-intensity beams.

\subsection{Distribution comparison and mismatch}
Let $f$ and $f'$ denote two centered Gaussian distributions with covariance matrices $\Sigma$ and $\Sigma'$, respectively.  To determine the distance between these distributions, we consider an arbitrary Gaussian kernel $k$ of the form:
\begin{equation}
k(X,X')=\exp\left(-\frac{1}{2}(X-X')^TS(X-X')\right), \label{GaussianS}
\end{equation}
where $S$ is any symmetric, positive definite matrix.  Such a matrix may always be decomposed as $S=A^TA$ for some matrix $A$.
The MMD between $f$ and $f'$ may be obtained using (\ref{MMDint}) as:
\begin{align}
\gamma_k^2(f,f')&=\det\left(I+2\Sigma_N\right)^{-1/2}+\det(I+2\Sigma_N')^{-1/2} \notag \\
&-2\det(I+\Sigma_N+\Sigma_N')^{-1/2}, \label{MMDgauss}
\end{align}
where $I$ is the identity matrix, and the normalized covariance matrices are:
\begin{equation}
\Sigma_N=A\Sigma A^T,\quad\quad \Sigma_N'=A\Sigma' A^T.
\end{equation}
This yields a large class of examples for benchmarking the computation of MMD in any dimension.

As a special case of (\ref{MMDgauss}), consider two Gaussian beams $f$ and $f'$ described on a 2D phase space $(x,p_x)$ with identical emittance $\epsilon$, with covariance matrices: 
\begin{equation}
\Sigma=\epsilon\begin{pmatrix}
\beta & -\alpha \\
-\alpha & \gamma
\end{pmatrix},\quad\quad 
\Sigma'=\epsilon\begin{pmatrix}
\beta' & -\alpha' \\
-\alpha' & \gamma'
\end{pmatrix}.
\end{equation}
Taking $S=\Sigma^{-1}$ in (\ref{GaussianS}) and computing the MMD between $f$ and $f'$ using (\ref{MMDgauss}) gives:
\begin{equation}
\gamma_k^2(f,f')=\frac{1}{3}-\frac{1}{\sqrt{5+4\zeta}}. \label{mismatch}
\end{equation}
Here the result is expressed in terms of the linear beam mismatch parameter $\zeta$, given by:
\begin{equation}
\zeta=\frac{1}{2}\left(\beta\gamma'-2\alpha\alpha'+\gamma\beta'\right),\quad \zeta\geq 1.
\end{equation}
The same result is obtained by taking $S=(\Sigma')^{-1}$ in (\ref{GaussianS}).  Note that (\ref{mismatch}) vanishes when $\zeta=1$, and increases monotonically with increasing mismatch $\zeta$.  Thus, when the MMD can be expressed in terms of the linear mismatch, the result behaves as expected.

In addition to detecting differences based on the second beam moments, the MMD detects differences in
the details of two distributions.  For a 4D example relevant to beams, consider a K-V distribution $f_{KV}$ and a (4D) Gaussian distribution $f_G$ with the
same second moments. Using (\ref{MMDft}) gives the exact result:
\begin{equation}
\gamma_k(f_{KV},f_{G})=\sqrt{\frac{1}{9}-\frac{1}{2e}+\frac{I_2(4)+I_3(4)}{2e^4}},
\end{equation}
where $I_n$ denotes the modified Bessel function of order $n$.
This corresponds to the numerical value $\gamma_k(f_{KV},f_{G})\approx 0.12863$.  For comparison, a numerical computation of $\gamma_k$ using (\ref{MMDfast}) from two sampled beams with $m=n=10^5$ particles and $L=10^4$ frequency samples gives an estimated value of 0.128.

\subsection{Detecting phase space correlations}
For a Gaussian distribution of any dimension, one may detect linear correlations among the various degrees of freedom by using (\ref{MMDgauss}) in (\ref{HCor}).  As an example, consider a Gaussian distribution on a 2D phase space $(q,p)$ with the covariance matrix:
\begin{equation}
\Sigma=\begin{pmatrix}
1 & r \\
r & 1 \\
\end{pmatrix},\quad f(q,p)=\frac{1}{2\pi a}e^{-(q^2+p^2-2rqp)/2a^2},
\end{equation}
where $a=\sqrt{1-r^2}$.
Using a Gaussian kernel of unit width, 
the correlation $\mathcal{R}_k$ between the variables $q$ and $p$ is given by:
\begin{equation}
\mathcal{R}_k^2=\frac{g(r)}{g(1)},\quad g(r)=\frac{1}{3}+\frac{1}{\sqrt{9-4r^2}}-\frac{2}{\sqrt{9-r^2}}.
\end{equation}
This result is expressed in terms of the standard linear correlation coefficient $r$.  Note that $\mathcal{R}_k$ increases monotonically from 0 to 1 as
$|r|$ increases from 0 to 1.

\begin{figure}
\begin{center}
\includegraphics[width=3.0in]{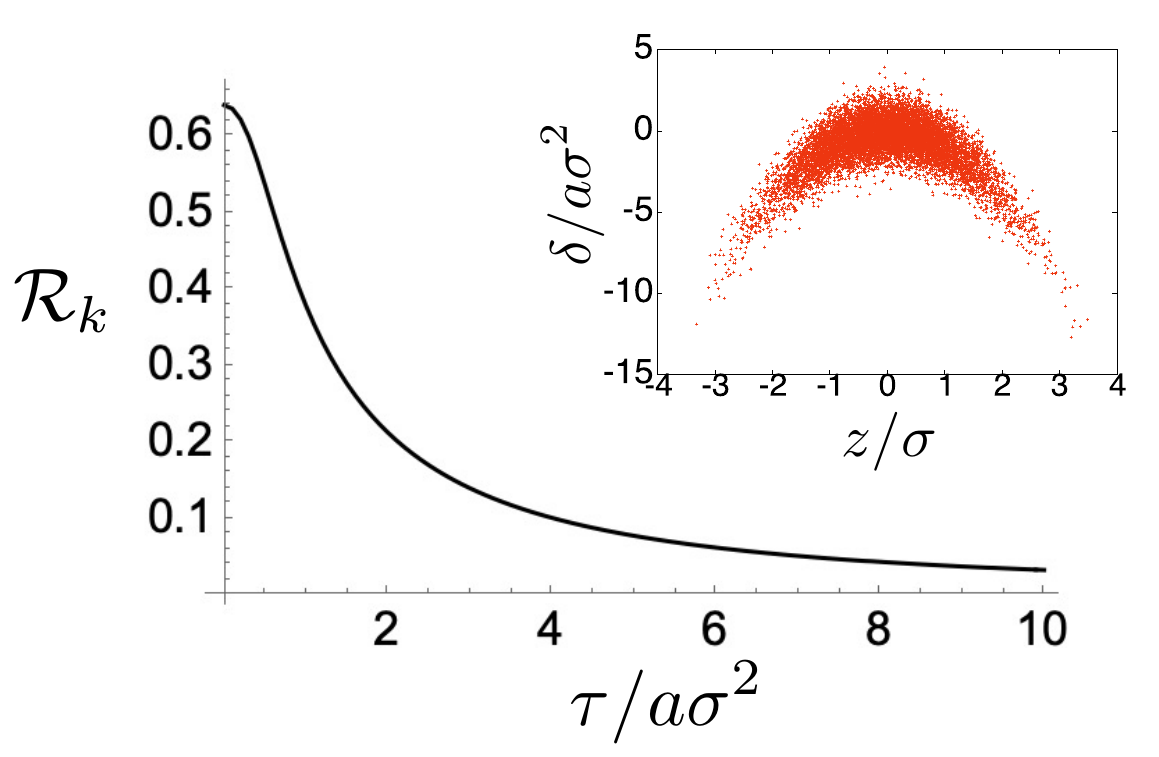}
\end{center}
\caption{Hilbert-Schmidt correlation $\mathcal{R}_k$ between variables $z$ and $\delta$ in the longitudinal phase space of a bunch with a quadratic correlation (\ref{quadf}).  (Inset) Sampled particles ($10^4$) for the case $\tau/a\sigma^2=1$, yielding the computed value
$\mathcal{R}_k\approx 0.375$.}  
\label{QuadCorFig}
\end{figure}

The quantity $\mathcal{R}_k$ is useful for detecting nonlinear correlations, even when the exact structure of the correlation is unknown.
Consider the case of a Gaussian beam with a quadratic correlation in the longitudinal phase space $(z,\delta)$:
\begin{equation}
f(z,\delta)=\frac{1}{2\pi\sigma\tau}e^{-z^2/{2\sigma^2}}e^{-(\delta+az^2)^2/2\tau^2}. \label{quadf}
\end{equation}
The linear correlation between $z$ and $\delta$ in (\ref{quadf}) vanishes, since one may verify that $\langle{z\delta\rangle}=0$.
Using the dimensionless variables $\bar{z}= z/\sigma$ and $\bar{\delta}=\delta/(a\sigma^2)$, and choosing a Gaussian kernel of width 1, one may evaluate the correlation $\mathcal{R}_k$ between $\bar{z}$ and $\bar{\delta}$.  The result is shown in Fig. \ref{QuadCorFig} as a function of the dimensionless parameter $\bar{\tau}=\tau/(a\sigma^2)$.  The result is independent of the sign of $a$.  We see that $\mathcal{R}_k$ becomes small as the quadratic coefficient $a$ becomes small or as the uncorrrelated energy spread $\tau$ becomes large, as one might expect.

\subsection{Testing for stationarity}
Often one must characterize the degree to which a given particle distribution remains stationary over many periods of dynamical evolution.  This can be quantified by computing the distance $\gamma_k(f_t,f_0)$ between the initial distribution $f_0$ and the distribution $f_t$ after $t$ periods.

For example, consider the 2D nonlinear symplectic map given by \cite{JMathPhys}:
\begin{equation}
\begin{pmatrix} q^f \\ p^f \end{pmatrix} = 
\begin{pmatrix}
\cos\phi& \sin\phi \\
-\sin\phi & \cos\phi
\end{pmatrix}
\begin{pmatrix}
q \\ p
\end{pmatrix},\quad \phi=\psi+\frac{\alpha}{2}\left(q^2+p^2\right), \label{toymap}
\end{equation}
where $\psi>0$ and $\alpha>0$ are constants.  This may be viewed as a simple model of a betatron phase advance in a single plane that increases linearly with the action $J=(q^2+p^2)/2$.

\begin{figure}
\begin{center}
\includegraphics[width=3.2in]{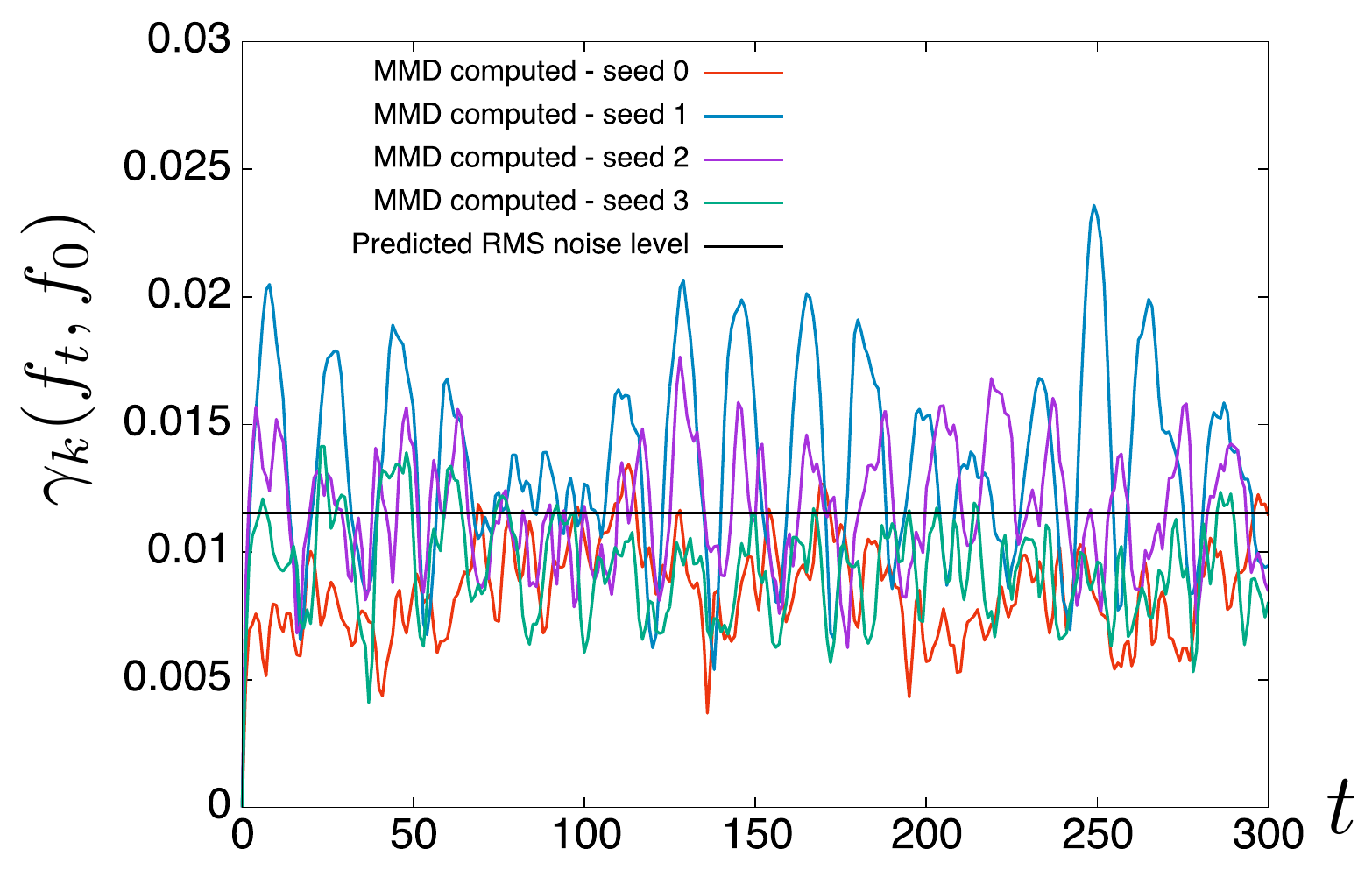}
\end{center}
\caption{Dynamics of a 2D beam with $n=10^4$ particles sampled from a matched Gaussian distribution (\ref{fGauss}) evolving under iteration of the map (\ref{toymap}).  The quantity $\gamma_k(f_t,f_0)$ is shown as a function of the iteration number $t$ for 4 distinct random seeds, showing that the distribution remains stationary.}  
\label{MMDtoInitial}
\end{figure}

A Gaussian distribution of the form:
\begin{equation}
f(q,p)=\frac{1}{2\pi\epsilon_0}e^{-(q^2+p^2)/2\epsilon_0} \label{fGauss}
\end{equation}
is an explicit function of the action $J$, and is therefore invariant under the map (\ref{toymap}).  Sampling $n=10^4$ particles from (\ref{fGauss}) and tracking them under iterates of the map (\ref{toymap}), we compare the distribution at each iteration with the initial distribution.  The result is shown in Fig. \ref{MMDtoInitial}.  

The MMD distance $\gamma_k(f_t,f_0)$ to the initial distribution is nonzero after the first iteration, but remains near $10^{-2}$ over the time interval observed.  Due to the finite number of particles, the value of $\gamma_k(f_t,f_0)$ experiences statistical fluctuations around the predicted rms value given by (\ref{MMDnoise}) (black curve).

\subsection{Relaxation to a stationary state}
If the distribution (\ref{fGauss}) is given an initial centroid offset $q\mapsto q+q_0$ with $q_0>0$, the beam will filament and converge weakly to a stationary equilibrium of the form \cite{Meller, JMathPhys}:
\begin{equation}
f_{eq}(q,p)=\frac{1}{2\pi\epsilon_0}e^{-(q^2+p^2+q_0^2)/2\epsilon_0}I_0\left(\frac{q_0}{\epsilon_0}\sqrt{q^2+p^2}\right). \label{toyeq}
\end{equation}
Using a Gaussian kernel of width $\sigma$ in expression (\ref{MMDft}), one may solve exactly for the time evolution of the MMD distance to equilibrium:
\begin{equation}
\gamma_k^2(f_t,f_{eq})=2s^2\sum_{n=1}^{\infty}\nu_ne^{-\nu_n(1+2d^2s^2\tau_n^2)}I_n(\nu_n), \label{exactMMDevolve}
\end{equation}
where $I_n$ denotes the modified Bessel function of order $n$ and 
\begin{equation}
\nu_n=\frac{d^2}{1+d^2s^2(1+\tau_n^2)},
\end{equation}
is given in terms of the dimensionless parameters:
\begin{equation}
\tau_n=nt\alpha\epsilon_0,\quad d^2=\frac{q_0^2}{2\epsilon_0},\quad s=\frac{\sigma}{q_0}.
\end{equation}
In particular, we see that for large $t$, (\ref{exactMMDevolve}) converges to zero as $\gamma_k\sim O(1/t^2)$.  

\begin{figure}
\begin{center}
\includegraphics[width=3.2in]{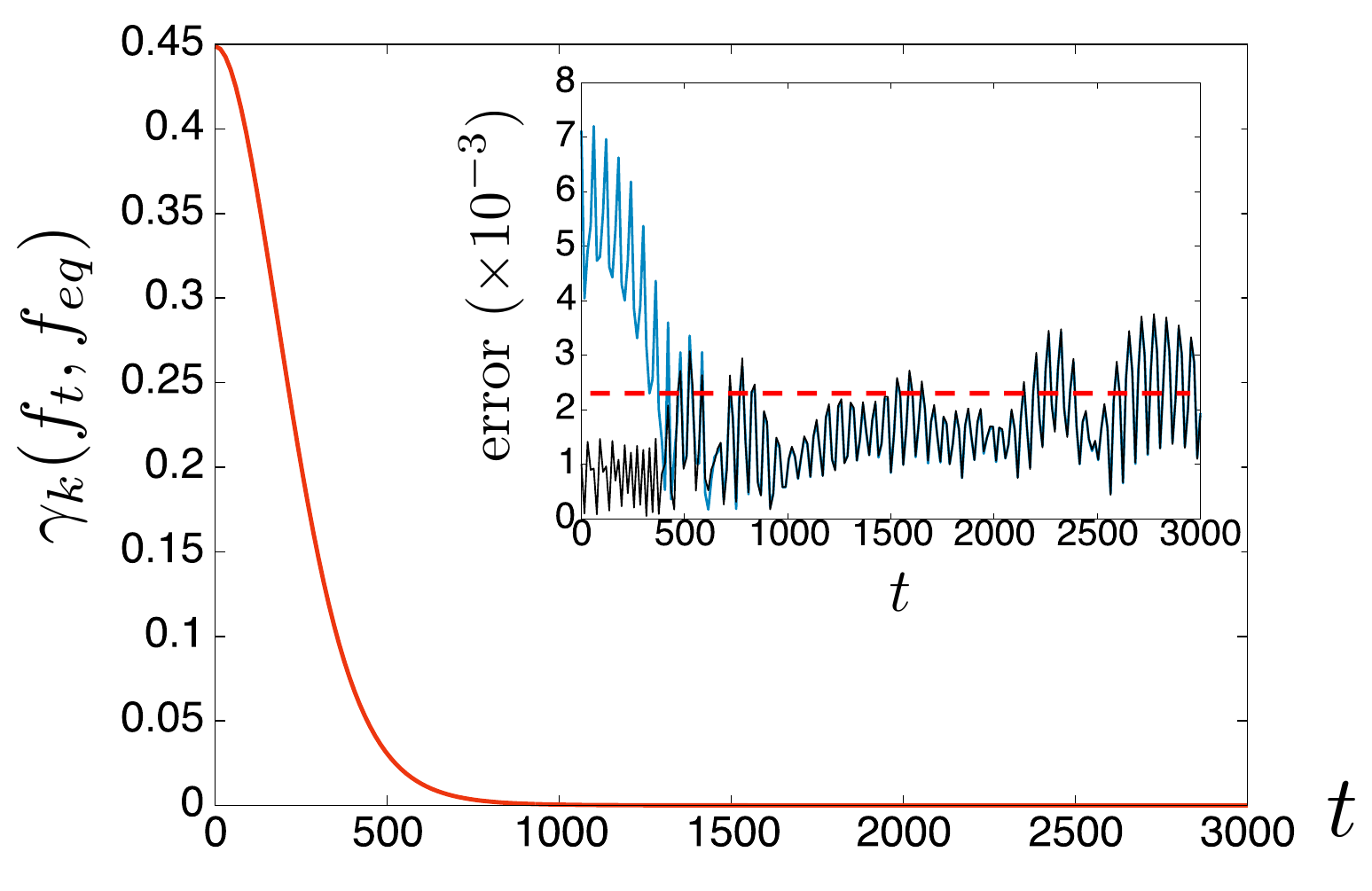}
\end{center}
\caption{Evolution of a distribution with $n=10^5$ particles sampled from (\ref{fGauss}) with initial offset $q\mapsto q+q_0$ under iteration of the map (\ref{toymap}). The distance of the distribution to the predicted equilibrium (\ref{toyeq}) after $t$ iterates is shown.  (Red curve) Analytical prediction (\ref{exactMMDevolve}).  (Black, inset) Absolute error in the numerical result obtained using (\ref{alt}).  (Blue, inset) Absolute error in the numerical result obtained using (\ref{MMDfast}). (Red, dashed) Prediction (\ref{MMDnoise}) of numerical noise using the distribution $f_{eq}$.}  
\label{MMDtoEq}
\end{figure}

Fig. \ref{MMDtoEq} illustrates the result obtained from tracking $10^5$ particles for the case $\epsilon_0=0.01$, $q_0=0.5$, $\psi=0.3$, $\alpha=0.1$.  We use a Gaussian kernel with $\sigma=1$.  The inset shows the difference from the prediction.  By $t=500$, the distribution has converged to equilibrium to within the resolution set by the particle noise.
This shows that MMD provides a diagnostic capable of measuring the dynamical relaxation of a beam to a stationary state.

\subsection{Mixing and decay of correlations}
Although the term ``mixing" is sometimes used to refer to any process involving filamentation and relaxation of the beam in phase space, we distinguish between regular mixing
(which is characteristic of nonlinear integrable systems) and mixing in the ergodic sense (which is characteristic of systems with widespread chaos).  Here, we refer to the latter meaning of the term, as it is formalized in ergodic theory \cite{Walters}.  (See Appendix B.)

A simple illustration of chaotic mixing behavior is given by the Arnold cat map, which 
is the 2D symplectic map given by:
\begin{equation}
\begin{pmatrix} q^f \\ p^f \end{pmatrix}=
\begin{pmatrix} 
2 & 1 \\
1 & 1
\end{pmatrix}
\begin{pmatrix} q \\ p \end{pmatrix} \mod 2\pi, \label{catmap}
\end{equation}
where we assume that $q$ and $p$ each have period $2\pi$.  The uniform density:
\begin{equation}
f(q,p)=\frac{1}{(2\pi)^2},\quad q,p\in [0,2\pi) \label{unifdist}
\end{equation}
is invariant under (\ref{catmap}) since the map is area-preserving.

\begin{figure}
\begin{center}
\includegraphics[width=3.2in]{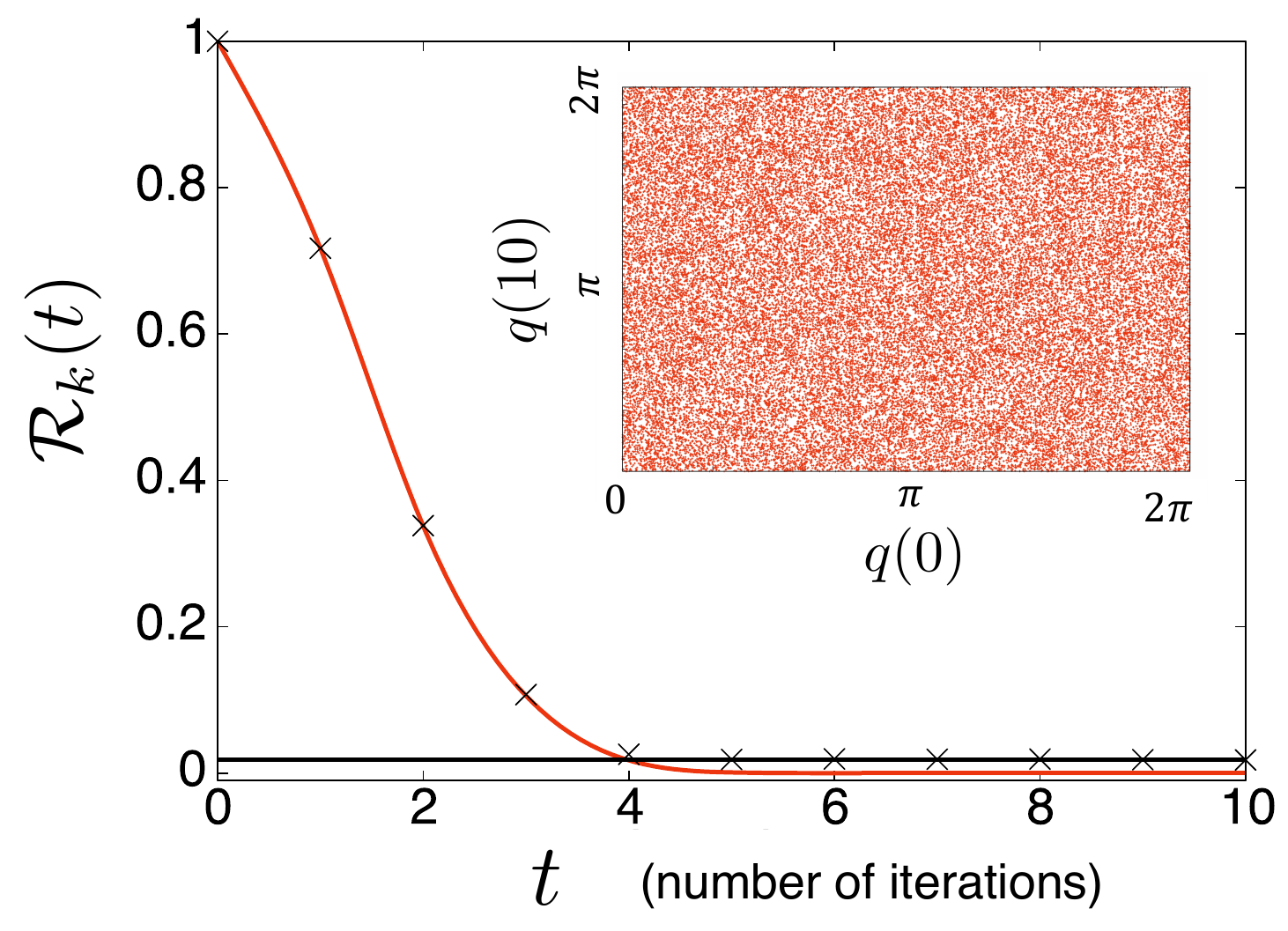}
\end{center}
\caption{Dynamics of a beam with $n=10^5$ particles sampled from the density (\ref{unifdist}) evolving under iteration of the map (\ref{catmap}).  The quantity $\mathcal{R}_k(t)$ is shown as a function of the iteration number $t$, illustrating the decay of correlations due to mixing.  The red curve shows the prediction (\ref{Rpred}), the black points are the results of simulation, and the black curve denotes the expected rms value due to noise (\ref{HSICnoise}).  (Inset) Plot of initial $q$ vs. final $q$ after 10 iterations, showing no visible correlation.}  
\label{CatFig}
\end{figure}

The Hilbert Schmidt correlation provides a quantitative measure of mixing and the resulting decay of correlations over time.  To illustrate this, we 
compute the value $\mathcal{R}_k(X(t),X(0))$, where $X(0)$ is the random variable denoting a particle's initial phase space coordinates, as sampled from (\ref{unifdist}), and $X(t)$ denotes the particle's phase space coordinates after $t$ iterations of the map (\ref{catmap}).  For simplicity, we denote this quantity by $\mathcal{R}_k(t)$.
In a periodic domain, it is appropriate to use a kernel that reflects the underlying domain periodicity. 
Using a Poisson kernel with parameter $\sigma$ in each dimension (Appendix A), we may compute the value of $\mathcal{R}_k(t)$ explicitly.
The value after $t$ iterations is given exactly by the sum:
\begin{equation}
\mathcal{R}_k(t)=\left(\frac{1-\sigma^2}{2\sigma}\right)\left[\sum_{\substack{n_1,n_2=-\infty \\ n_1\neq 0,n_2\neq 0}}^{\infty}\sigma^{C(n_1,n_2,t)}\right]^{1/2}, \label{Rpred}
\end{equation}
where the exponent is $(t=1,2,3,\ldots)$:
\begin{align}
C(n_1,&n_2,t)=|n_1|+|n_2| \notag \\
&+\left|F_{2t+1}n_1+F_{2t}n_2\right|+|F_{2t}n_1+F_{2t-1}n_2|
\end{align}
given in terms of the usual Fibonacci sequence:
\begin{equation}
F_1=F_2=1,\quad F_t=F_{t-1}+F_{t-2}.
\end{equation}
Fig. \ref{CatFig} shows the decay of the quantity $\mathcal{R}_k(t)$ as a function of the iteration number $t$ for the case $\sigma=1/2$.  Because the prediction (\ref{Rpred}) is only defined for nonnegative integer values of $t$, the red curve shown connects these values using smooth interpolation.  Note that mixing for the map (\ref{catmap}) is very rapid.  After only 4-5 iterations, the correlations between the initial and final phase space coordinates are at or below the level expected due to numerical particle noise, as given by (\ref{HSICnoise}) and indicated by the black line.  The inset shows that no correlation is visible by eye between the initial and final coordinates.

\section{Application to High-Intensity Beams}\label{sec:dynamic}

\subsection{Beam in a constant focusing channel}
As our first example with space charge, we consider a long (unbunched) intense beam in a constant focusing channel that is initialized in a stationary thermal equilibrium, so the 4D beam distribution takes the form:
\begin{equation}
f_0(x,p_x,y,p_y)\propto e^{-H(x,p_x,y,p_y)/kT}, \label{thermal}
\end{equation}
where $H$ denotes the self-consistent Hamiltonian:
\begin{equation}
H=\frac{1}{2}(p_x^2+p_y^2)+\frac{1}{2}\Omega^2(x^2+y^2)+\frac{q\phi(x,y,s)}{\beta^2\gamma^3mc^2}, \label{HamCF}
\end{equation}
and $\phi$ is a solution of the 2D Poisson equation:
\begin{equation}
\nabla_{\perp}^2\phi=-\frac{\lambda}{\epsilon_0}\int f_0(x,p_x,y,p_y)dp_xdp_y.
\end{equation}
Here $\lambda$ denotes the charge per unit length, and $\epsilon_0$ denotes the vacuum permittivity.

For simulation, we consider a proton beam with a kinetic energy of 200 MeV and a beam current of 20 A in an external focusing of strength $\Omega=0.628$ $m^{-1}$ (corresponding to a 2.7 T solenoid field).  The temperature $kT$ is chosen to yield the initial emittance $\epsilon_{x,rms}=\epsilon_{y,rms}=1.24$ $\mu$m.  The tune depression due to space charge is then given by $\nu/\nu_0\approx 0.55$.  

\begin{figure}
\begin{center}
\includegraphics[width=3.2in]{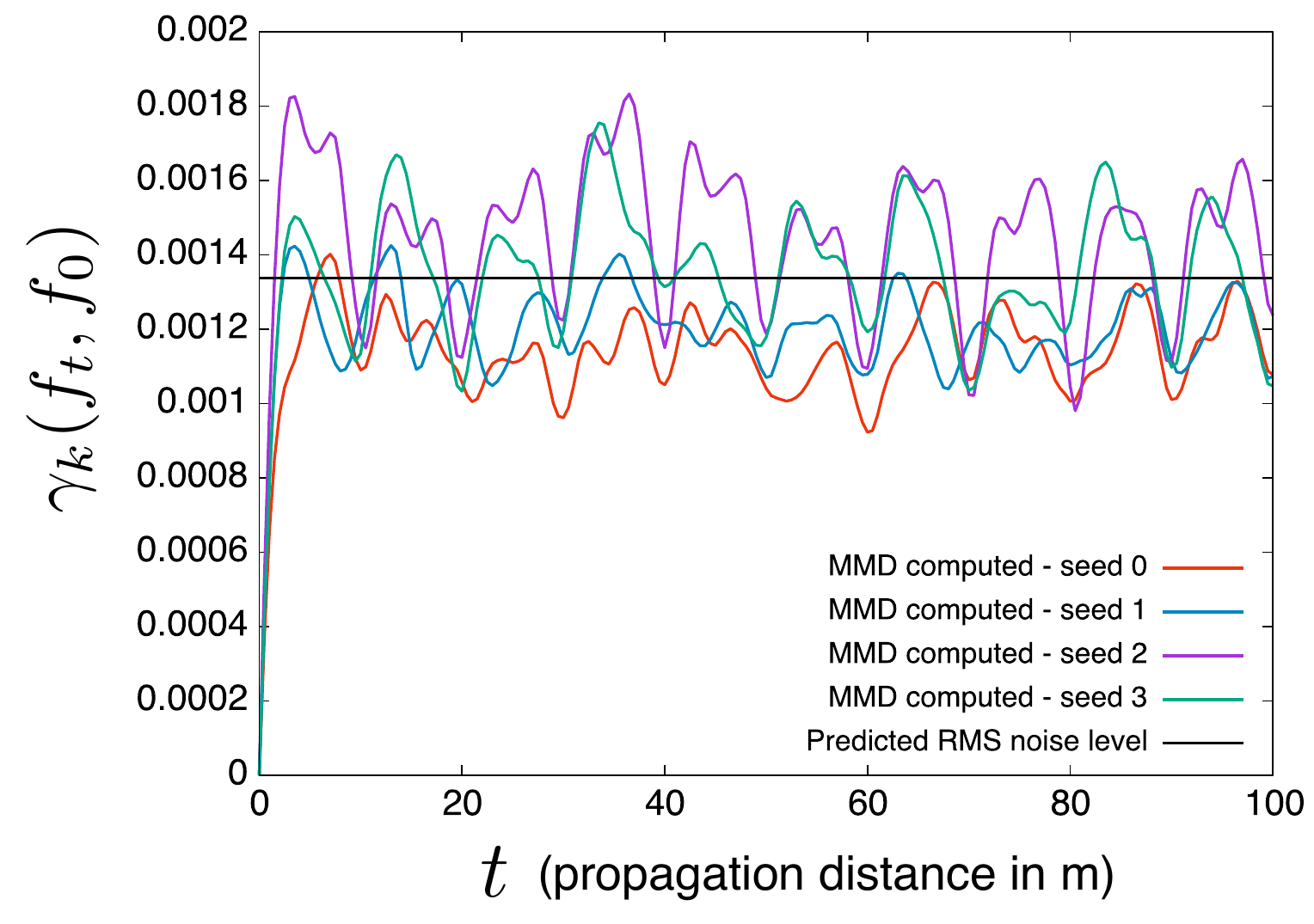}
\end{center}
\caption{Test of stationarity for an unbunched (4D) beam with $n=10^6$ particles sampled from a thermal equilibrium distribution (\ref{thermal}) propagating in a linear constant focusing channel.  The quantity $\gamma_k(f_t,f_0)$ is shown as a function of propagation distance $t$ for 4 distinct random seeds.}  
\label{CFChannel}
\end{figure}

Fig. \ref{CFChannel} illustrates the MMD distance between the initial distribution and the distribution at time $t$, for four distinct random seeds.  To compute (\ref{MMDfast}), a Gaussian kernel was used.  The kernel width parameter $\sigma$ associated with each phase space dimension was matched to the corresponding rms width of the distribution (\ref{thermal}).  For each random seed, self-consistent tracking with transverse (2D) space charge using $n=10^6$ particles was performed using the symplectic gridless spectral solver described in \cite{Qiang}.  Notice that the distribution remains stationary to within the level expected due to particle noise (black line).  Compare this dynamical behavior to that shown in Fig. \ref{MMDtoInitial}.

\begin{figure}
\begin{center}
\includegraphics[width=3.0in]{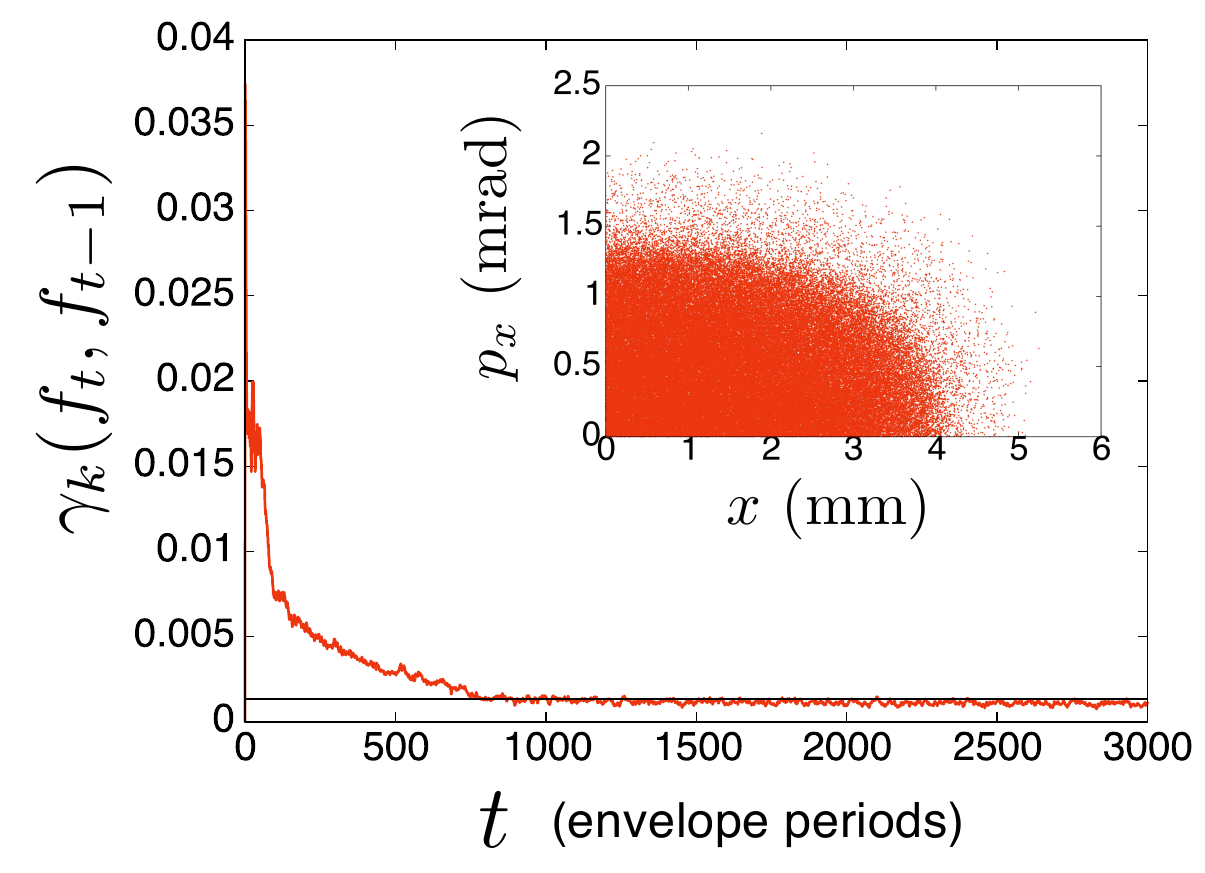}
\end{center}
\caption{Collisionless relaxation of an unbunched (4D) beam with $n=10^6$ particles sampled from the distribution (\ref{Levdist}) in a linear constant focusing channel (\ref{HamCF}), where $t$ denotes the number of linearized envelope periods.  The MMD between the distribution on successive periods $\gamma_k(f_t,f_{t-1})$ decays to the level of noise as the beam relaxes to a stationary state.  (Inset) The final particle distribution (one quadrant).}  
\label{CFMismatched}
\end{figure}

As a second example, consider a proton beam with the same energy and emittance as above.  However, instead of the stationary distribution (\ref{thermal}), we use the initial distribution \cite{Levin1}:
\begin{equation}
f_0(x,p_x,y,p_y)=\frac{1}{\pi^2r_m^2p_m^2}\Theta(r_m-r)\Theta(p_m-p) \label{Levdist}
\end{equation}
where $r=\sqrt{x^2+y^2}$, $p=\sqrt{p_x^2+p_y^2}$, and $\Theta$ denotes the unit step function.  The distribution (\ref{Levdist}) is not stationary, but to minimize fluctuations of the rms beam size, we match the beam in an rms sense by setting $p_m^2=\Omega^2r_m^2-K_{pv}$, where $K_{pv}$ denotes the generalized beam perveance.  In the absence of precise knowledge of the final equilibrium state, we compare the particle distribution at successive time intervals separated by $\Delta t = L$, where $L=6.20$ m denotes the period of linearized envelope oscillations about the equilibrium beam size.  The result is shown in Fig. \ref{CFMismatched}.  On the time scale shown, the beam appears to undergo relaxation toward a final distribution containing a small but visible low-density halo (inset).  This fast relaxation appears to be a property of the collisionless Vlasov-Poisson system, and it is to be distinguished from slow relaxation due to collisional effects \cite{LongRange}, which are not included in the numerical model.

The system described in this section cannot be mixing due to the existence of invariants of motion (for example, the angular momentum).  Indeed, the computed quantity $\mathcal{R}_k(t)$ converges rapidly to a nonzero value.  This behavior will be examined using a more complex example in the following section.

\subsection{Beam in a periodic FODO lattice}
\begin{figure}
\begin{center}
\includegraphics[width=3.0in]{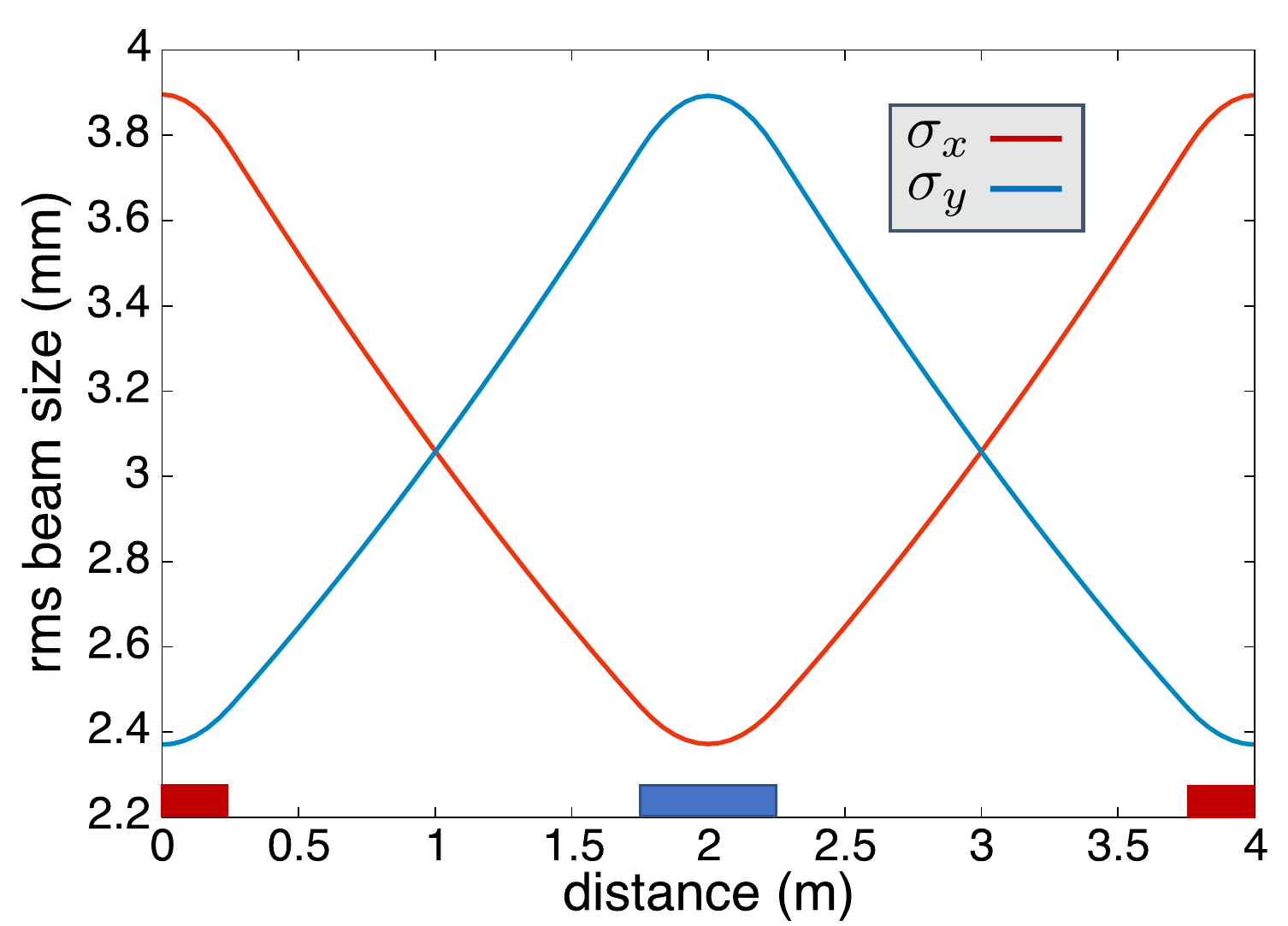}
\end{center}
\caption{Matched (K-V) beam envelopes for a 10 A proton beam at 200 MeV in the FODO cell used in Section VI.b.  Red rectangle - focusing quad.  Blue rectangle - defocusing quad.}  
\label{FODOcell}
\end{figure}

As a simple application to an intense beam in a periodic focusing structure, we consider a lattice consisting of a single FODO cell (Fig. \ref{FODOcell}).
We consider a proton beam with a kinetic energy of 200 MeV and a beam current of 10 A with an initial emittance of $\epsilon_{x,rms}=\epsilon_{y,rms}=1$ $\mu$m (unnormalized).  The initial distribution is Gaussian of the form:
\begin{equation}
f_0(x,p_x,y,p_y)=\frac{1}{(2\pi)^2\epsilon_{x,rms}\epsilon_{y,rms}}e^{-\frac{1}{2}X^T\Sigma^{-1}X}, \label{GaussFODO}
\end{equation}
where $X=(x,p_x,y,p_y)$ and $\Sigma$ denotes the covariance matrix.
Although the second beam moments are chosen so that the beam is matched in an rms-sense, the distribution is not matched in detail.  Fig. \ref{FODOcell} shows the evolution of the matched beam envelopes over a single period.  The zero-current phase advance is $60.1^{\circ}$ per cell, while the 10 A phase advance is $25.9^{\circ}$ per cell, so that space charge plays a significant role.

\begin{figure}
\begin{center}
\includegraphics[width=2.8in]{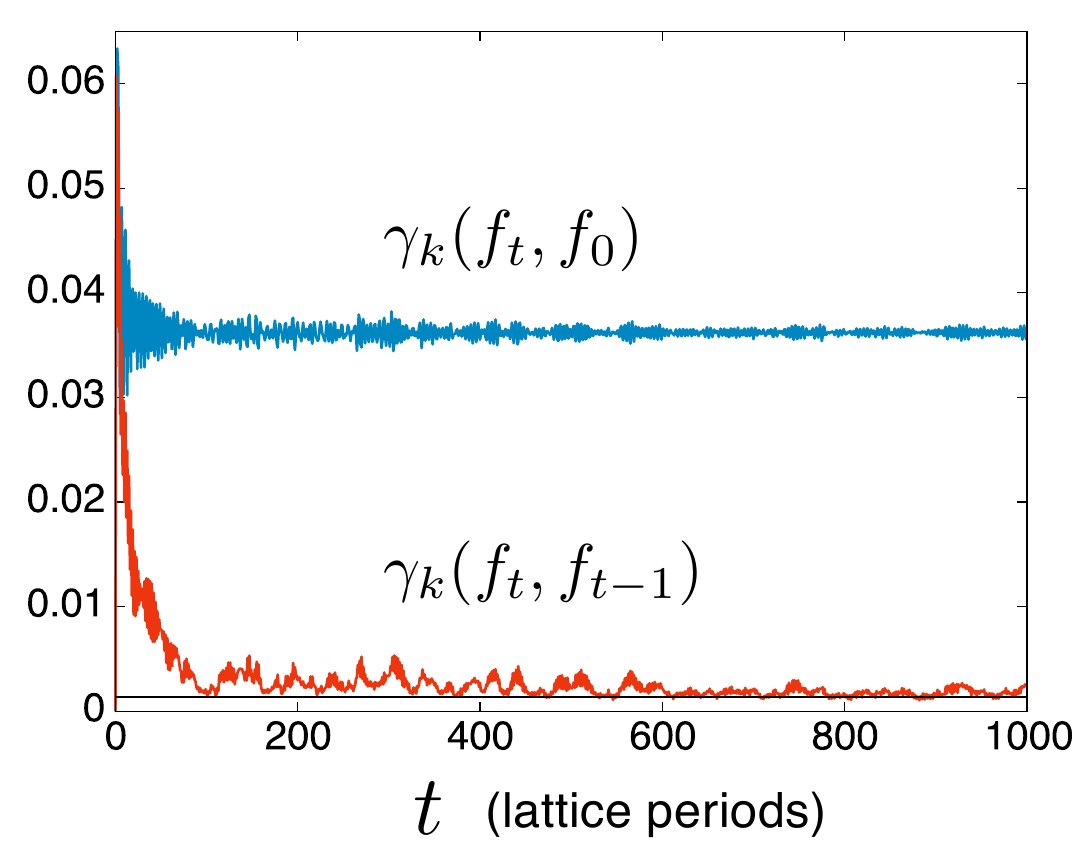}
\end{center}
\caption{Dynamics of an unbunched (4D) beam with $n=10^6$ particles sampled from the distribution (\ref{GaussFODO}) in the FODO channel shown in Fig. \ref{FODOcell}.  The MMD between the distribution on successive lattice periods $\gamma_k(f_t,f_{t-1})$ decays to near (but remains slightly above) the level of noise (black line).  The MMD to the initial distribution $\gamma_k(f_t,f_0)$ is largely unchanged after the first 100 periods.}  
\label{FODOmmd}
\end{figure}

Fig. \ref{FODOmmd} illustrates the MMD distance between the initial distribution and the distribution after $t$ periods (blue), together with the MMD distance between successive periods (red).  To compute (\ref{MMDfast}), a Gaussian kernel was used whose width along each dimension matches the initial rms beam size in that dimension.  In each case, self-consistent tracking with transverse (2D) space charge using $n=10^6$ particles was performed using the algorithm described in \cite{Qiang}.  The distribution relaxes quickly over the first 100 periods, but fluctuations above the noise level persist on a much longer time scale.  

\begin{figure}
\begin{center}
\includegraphics[width=2.8in]{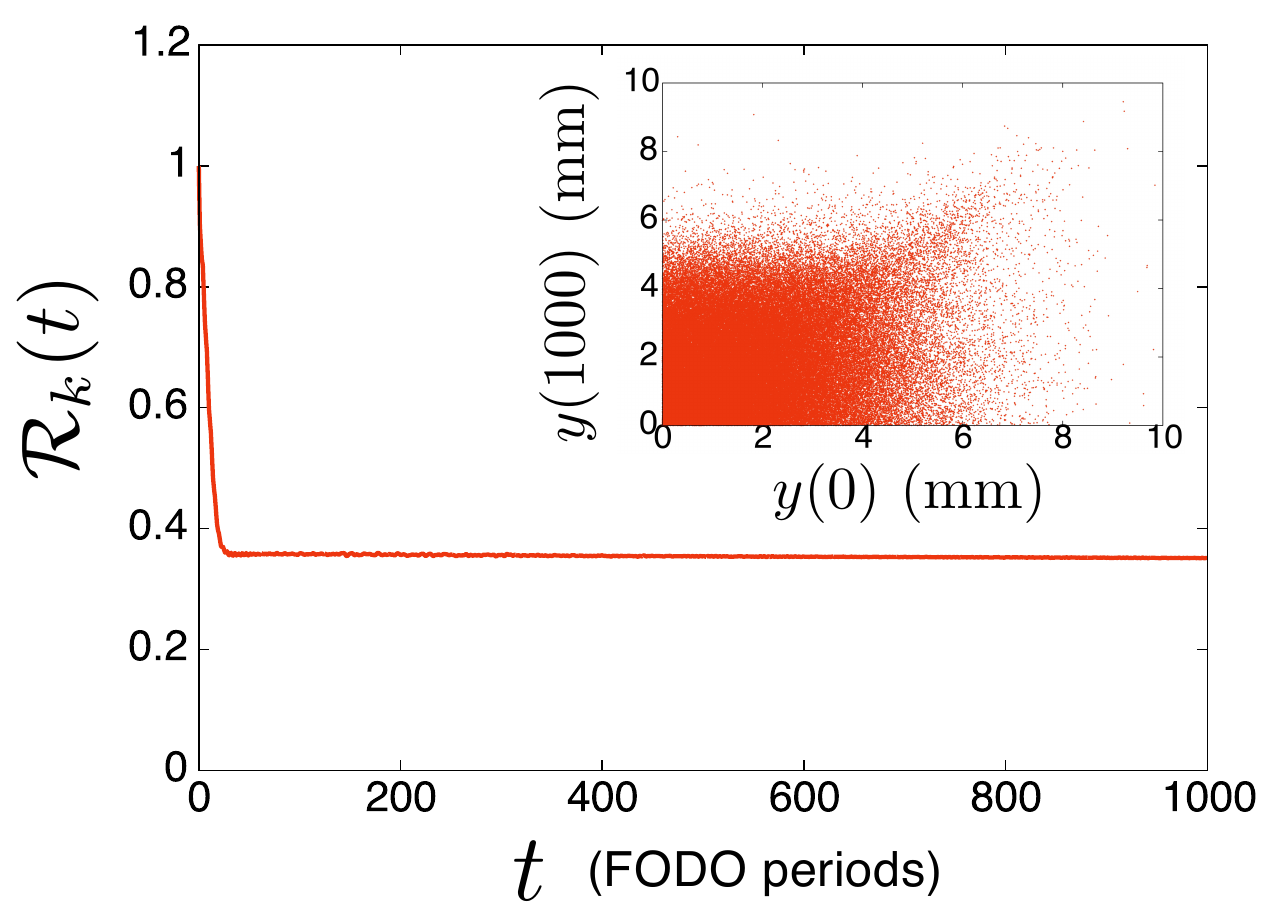}
\end{center}
\caption{Dynamics of an unbunched (4D) beam with $n=10^6$ particles sampled from the distribution (\ref{GaussFODO}) in the FODO channel shown in Fig. \ref{FODOcell}.  The correlation $\mathcal{R}_k(t)$ between the distribution after $t$ periods and the initial distribution at $t=0$ is shown.  Correlations appear to persist indefinitely.  (Inset) Plot of initial $y$ vs. final $y$ after 1K periods, showing that the correlations are not easily visible in low-dimensional projections.}  
\label{FODOcorr}
\end{figure}

Fig. \ref{FODOcorr} shows the correlation $\mathcal{R}_k(t)$ between the distribution at time $t$ and the distribution at $t=0$.  Note that $\mathcal{R}_k(t)$ converges to a fixed nonzero value within just a few periods, and then remains constant.   
This indicates that particle coordinates remain correlated with their initial values indefinitely, and that the dynamics is not mixing. 
This generally suggests the existence of invariants of motion in some regions of the phase space.  A 2D plot of $y(0)$ versus $y(1000)$ is also shown.  Note that the correlations are not easily visible.  In fact, if $\mathcal{R}_k$ is computed using only the initial and final values of $y$, neglecting all other coordinates, then the corresponding value is 0.08.  This shows that $\mathcal{R}_k$ can quantify correlations present in higher dimensions that are not easily visualized by viewing 2D projections.

\subsection{Treatment of beam loss}
In the presence of beam loss, it may be of interest to study the dynamics of the beam on a bounded subregion $E$ of the phase space (e.g., defined by the vacuum chamber or by the dynamic aperture).  For example, one may study the relaxation of the distribution defined by those particles that remain indefinitely within the region $E$.
In this case, the beam distribution function $f_t$ satisfies:
\begin{equation}
\int_E f_t(X)dX=\chi_t,\quad 0\leq \chi_t\leq 1,
\end{equation}
where $\chi_t$ denotes the (possibly time-dependent) fraction of beam particles that lie within the region $E$.

The formalism of the previous section may be modified to treat this case as follows.
 \begin{itemize}
 \item{In the calculation of (\ref{MMDsample}-\ref{MMDfast}) and (\ref{HSICquad}-\ref{RFF}), sum only over those particles that lie within the desired region $E$.  This is equivalent to modifying the kernel $k$ by setting $k(X,X')=0$ if $X\notin E$ or $X'\notin E$.}
 \item{In the calculation of (\ref{MMDsample}-\ref{MMDfast}) and (\ref{HSICquad}-\ref{RFF}), replace the weight coefficients $1/m$ by particle weights $w_j$ with $\sum_{j=1}^mw_j=\chi_t$, and similarly for $1/n$.}
 \end{itemize}
 One may verify that many of the desired mathematical properties of the MMD and HSICor still hold for these modified statistics.
  
\section{Conclusions}\label{sec:concl}
Modeling beams in the presence of high intensity space charge, collective instabilities, or strong nonlinear focusing can result in dynamical processes that are difficult to characterize using typical numerical diagnostics based on second-order moments.  We have introduced two numerical diagnostics originating in the ML literature with highly desirable mathematical properties that are straightforward to implement in parallel particle-based simulation codes.  The first is a measure of statistical distance known as the Maximum Mean Discrepancy ($\gamma_k$), which serves as a measure of similarity between two particle ensembles.  The second is a measure of statistical dependence or correlation between random variables known as the Hilbert Schmidt Correlation ($\mathcal{R}_k$).  These quantities are useful for a variety of applications, including:  matching a beam into a periodic transport system, numerical benchmarking, detecting possibly nonlinear phase space correlations, characterizing relaxation to a (quasi-)stationary state, and characterizing mixing or decay of correlations within the beam.  

It is important to note that the quantitative results obtained will depend on the choice of the kernel $k$.  On one hand, this kernel-dependence may be viewed as a disadvantage of the diagnostics described here.  On the other hand, one may view the kernel as a natural way to parameterize a large family of possible diagnostics, all of which correctly capture the same underlying physical processes.  (This is a consequence of the mathematical properties described in Sections III-IV.)  In numerical experiments the observed dynamical evolution of $\gamma_k$ or $\mathcal{R}_k$ was largely independent of the choice of kernel, although this remains an active area of investigation.  An alternative and parameter-free statistical distance with similar mathematical properties is the Wasserstein distance $W_p$ (Section II.B).  In the future, the authors plan to investigate the feasibility of using efficient approximations  to $W_p$ \cite{WassSmooth} for beam dynamics applications.

The diagnostics described here are well-suited to applications involving large simulation ensembles.
For example, quantities involving $\gamma_k$ or $\mathcal{R}_k$ may be used as objectives for accelerator design optimization or for training machine learning models that require detailed information about the beam distribution function.  This raises the possibility of tailoring the final beam phase space density by using large-scale automated machine tuning.

Finally, although we have focused on the case of charged particle beams, it is clear that these tools can be applied without change to kinetic simulations of other many-body systems such as plasmas or gravitational systems, which rely on the tracking of large particle ensembles.

\section{Acknowledgments}
This work was supported by the Director, Office of Science of the U.S. Department of Energy under Contract No. DE-AC02-05CH11231, and made use of computer resources at the National Energy Research Scientific Computing Center.   The authors acknowledge support from the U.S. DOE Early Career Research Program under the Office of High Energy Physics.  
 \\
\section*{Appendix A:  Commonly-used kernels}
This Appendix lists several of the kernels $k$ most commonly used for discrimination testing and independence testing in ML.  These kernels are all translation-invariant and normalized as in (\ref{knormal}), so that $k(X,X)=1$ for all $X$.  All the kernels listed here have the property that the quantity $\gamma_k$ in (\ref{MMDint}) satisfies the metric conditions (i-iv) and captures weak convergence, as described in Section II.

Gaussian kernel $(\sigma>0)$ of dimension $d$:
\begin{subequations}\label{kGaussian}
\begin{align}
&k_{\rm Gaussian}(X,X')=e^{-|X-X'|^2/2\sigma^2}, \\
&\Lambda_{\rm Gaussian}(\omega)=\frac{e^{-\sigma^2|\omega|^2/2}}{(2\pi\sigma^2)^{d/2}}.
\end{align}
\end{subequations}

Laplace kernel $(\sigma>0)$ of dimension $d$:
\begin{subequations}\label{kLaplace}
\begin{align}
&k_{\rm Laplace}(X,X')=e^{-|X-X'|/\sigma}, \\
&\Lambda_{\rm Laplace}(\omega)=\frac{\sigma^d}{\pi^{(d+1)/2}}\frac{\Gamma\left(\frac{d+1}{2}\right)}{( 1+\sigma^2\omega^2 )^{(d+1)/2}},
\end{align}
\end{subequations}
where $\Gamma$ denotes the gamma function.

Mat\'ern kernel ($\nu>0$, $\sigma>0$) of dimension $d$:
\begin{subequations}\label{kMatern}
\begin{align}
&k_{\rm Matern}(X,X')=\frac{2^{1-\nu}}{\Gamma(\nu)}\zeta^{\nu}K_{\nu}\left(\zeta\right), \\
&\Lambda_{\rm Matern}(\omega)=\frac{2^{s}\sigma^d\nu^{\nu}\Gamma(s)}{(2\pi)^{d/2}\Gamma(\nu)}\left({2\nu}+\sigma^2\left|\omega\right|^2\right)^{-s},
\end{align}
\end{subequations}
where $K_{\nu}$ is the modified Bessel function of order $\nu$, and we abbreviate:
\begin{equation}
\zeta=\frac{\sqrt{2\nu}}{\sigma}|X-X'|,\quad\quad s=\nu+\frac{d}{2}.
\end{equation}

Poisson kernel ($0<\sigma<1$) of dimension 1:
\begin{subequations}\label{kPoisson}
\begin{align}
&k_{\rm Poisson}(X,X')=\frac{(1-\sigma)^2}{1-2\sigma\cos(X-X')+\sigma^2}, \\
&\Lambda_{\rm Poisson}(\omega)=\sum_{n=-\infty}^{\infty}\left(\frac{1-\sigma}{1+\sigma}\right)\sigma^{|n|}\delta(\omega-n).
\end{align}
\end{subequations}

Another class of kernels $k_{\rm Wendland}$ (of dimension $d$) is constructed by using a polynomial with compact support in the separation distance $r=|X-X'|/\sigma$, where $\sigma>0$.  See  \cite{Wendland, Wendland2} for a detailed description of these.  An example for $d=1$ is given by:
\begin{subequations}\label{kWendland}
\begin{align}
&k_{\rm Wend}(X,X')=(1-r)^3(1+3r),\quad\quad r<1, \\
&\Lambda_{\rm Wend}(\omega)=\frac{24\sigma}{\pi}\left\{\frac{2+\cos\omega\sigma}{(\omega\sigma)^4}-\frac{3\sin\omega\sigma}{(\omega\sigma)^5}\right\}.
\end{align}
\end{subequations}

When working with a Gaussian kernel, an orthonormal basis for the RKHS is given by $\{e_n\}_{n=0}^{\infty}$ where \cite{Gaussian}:
\begin{equation}
e_n(X)=\sqrt{\frac{1}{\sigma^{2n}n!}}e^{-X^2/2\sigma^2}X^{n}.
\end{equation}

When working with the Poisson kernel, a (complex) orthonormal basis for the RKHS is given by $\{e_n\}_{n=-\infty}^{\infty}$ with:
\begin{equation}
e_n(X)=\sqrt{\frac{1-\sigma}{1+\sigma}}\sigma^{|n|/2}e^{inX}.
\end{equation}
A corresponding real basis is easily constructed.

We can construct a kernel $k$ of higher dimension using kernels $k^{(j)}$ $(j=1,\ldots,d)$ of lower dimension.  For example, if we write $X=(X_1,\ldots,X_d)$ and $X'=(X_1',\ldots,X_d')$, then
\begin{equation}
k(X,X')=\prod_{j=1}^dk^{(j)}(X_j,X_j').
\end{equation}
If the kernels $k^{(j)}$ are translation-invariant with spectral densities $\Lambda^{(j)}$, then setting $\omega=(\omega_1,\ldots,\omega_d)$ we have:
\begin{equation}
\Lambda(\omega)=\prod_{j=1}^d\Lambda^{(j)}(\omega_j).
\end{equation}
If the RKHS associated with the kernel $k^{(j)}$ has basis $e_n^{(j)}$, then the RKHS associated with $k$ has basis 
\begin{equation}
e_n(X)=\prod_{j=1}^d e^{(j)}_{n_j}(X_j),
\end{equation}
where $n=(n_1,\ldots,n_d)$ ranges over all possible indices.

\section*{Appendix B:  Definition of mixing}
Let $\mathcal{M}$ be a map, and let $f$ denote a probability density that is invariant under the map, in the sense that:
\begin{equation}
f(\mathcal{M}^{-1}(X))=f(X).
\end{equation}
Then the map $\mathcal{M}$ is mixing with respect to the invariant density $f$ if for any two sets $A$ and $B$ \cite{MixDef}:
\begin{equation}
\lim_{t\rightarrow\infty}P(\mathcal{M}^{-t}(A)\cap B)=P(A)P(B), \label{mixing}
\end{equation}
where 
\begin{equation}
P(A)=\int_A f(X)dX
\end{equation}
denotes the probability that a point lies in $A$.
Informally, for any sets $A$ and $B$, the sequence of sets $\mathcal{M}^{-t}(A)$ becomes asymptotically independent of $B$ as $t\rightarrow\infty$ \cite{Walters}.

There are many equivalent formulations of the condition (\ref{mixing}).  For our purposes, it is enough to know that a map $\mathcal{M}$ is mixing with respect to
a density $f$ if and only if $\mathcal{R}_k(t)\rightarrow 0$ as $t\rightarrow\infty$, where $\mathcal{R}_k(t)$ denotes the correlation between the particle coordinates sampled from $f$ at $t=0$ and the corresponding particle coordinates at later time $t$.  (See Section IV.)  This holds for all kernels $k$ satisfying the desirable properties described in Section II.C.


\begin{thebibliography}{9} 
\bibitem{Kandrup}
H. Kandrup {\it el al}, ``Chaotic collisionless evolution in galaxies and charged-particle beams," Ann. N. Y. Acad. Sci. {\bf 1045}, 12 (2005).
\bibitem{Bohn1}
C. Bohn, ``Chaotic dynamics in charged-particle beams:  possible analogs of galactic evolution," Ann. N. Y. Acad. Sci. {\bf 1045}, 34 (2005).
\bibitem{UMER}
D. Stratakis {\it et al}, ``Experimental and numerical study of phase mixing of an intense beam," Phys. Rev. ST Accel. Beams {\bf 12}, 064201 (2009).
\bibitem{LongRange}
Y. Levin {\it et al}, ``Nonequilibrium statistical mechanics of systems with long-range interactions," Phys. Rep. {\bf 535}, 1-60 (2014).
\bibitem{CoulombR}
B. B. Kadomtsev and O. P. Pogutse, 	``Collisionless relaxation in systems with Coulomb interactions," Phys. Rev. Lett. {\bf 25}, 1155 (1970).
\bibitem{Levin1}
Y. Levin, R. Pakter, and T. Teles, ``Collisionless Relaxation in Non-Neutral Plasmas," Phys. Rev. Lett. {\bf 100}, 040604 (2008).
\bibitem{MMDdef}
A. Gretton {\it et al}, ``A Kernel Two-Sample Test", Journal of Machine Learning Research {\bf 13}, 723-773 (2012).  
\bibitem{RKHSsurvey}
T. Hofmann, B. Scholkopf, and A. J. Smola, ``Kernel Methods in Machine Learning," The Annals of Statistics {\bf 36}, 1171-1220 (2008).
\bibitem{RKHStext}
V. Paulsen and M. Raghupathi, {\it An Introduction to the Theory of Reproducing Kernel Hilbert Spaces}, Cambridge University Press, 2016.
\bibitem{HSIC}
A. Gretton {\it et al}, ``Measuring Statistical Dependence with Hilbert-Schmidt Norms", Technical report no. 140, Max Planck Institute for Biological Cybernetics (2005),
\url{http://www.gatsby.ucl.ac.uk/~gretton/papers/GreHerSmoBouSch05a.pdf}.
\bibitem{DensityDef}
 By a ``probability density" we mean, more generally, a probability measure defined on the $\sigma$-algebra of Borel subsets of the phase space $\mathbb{R}^d$.  Such a measure may not possess a density in the strict sense of a function in $L^1(\mathbb{R}^d)$.  For example, our ``probability density" here is allowed to contain Dirac delta functions.
\bibitem{Dudley}
R. M. Dudley, {\it Real Analysis and Probability}, Cambridge Studies in Advanced Mathematics 74, Cambridge University Press, 2002.
\bibitem{Billingsley}
P. Billingsley, {\it Convergence of Probability Measures}, 2nd ed, Wiley, NY, 1999.
 \bibitem{Statdist}
M. Markatou, D. Karlis, and Y. Ding, ``Distance-Based Statistical Inference," Annu. Rev. Stat. Appl. {\bf 8}, 301-27 (2021).
\bibitem{KLdef}
S. Kullback and R. Leibler, ``On Information and Sufficiency," Annals of Mathematical Statistics {\bf 22}, 79-86 (1951).
\bibitem{KLbook}
S. Kullback, {\it Information Theory and Statistics}, Dover Publications, Inc., Mineola, NY (1968).
\bibitem{KLapp}
C. Granero-Belinchon, S. Roux, and N. Garnier, ``Kullback-Leibler divergence measure of intermittency: Application to turbulence," Phys. Rev. E {\bf 97}, 013107 (2018).
\bibitem{KLest}
Q. Wang {\it et al}, ``Divergence Estimation for Multidimensional Densities via $k$-Nearest-Neighbor Distances", IEEE Transactions on Information Theory {\bf 55}, 2392 (2009).
\bibitem{note1}
Strictly speaking, the infimum of this quantity must be taken over all ``joint" probability measures defined on the Borel subsets of $\mathbb{R}^d\times\mathbb{R}^d$ whose marginals have densities
$f$ and $g$, respectively.  See comment \cite{DensityDef}.
\bibitem{Wassdef1}
L. V. Kantorovich, ``On the Translocation of Masses", Dokl. Akad. Nauk USSR {\bf 37}, pp. 227-229 (1942).  [English translation:  Journal of Mathematical Sciences {\bf 133}, 1381-1382 (2006)].
\bibitem{Wassdef2}
L. N. Wasserstein, ``Markov Processes Over Denumerable Products of Spaces Describing Large Systems of Automata," Probl. Inform. Transmission {\bf 5}, pp. 47-52 (1969).
\bibitem{OptTransp}
C. Villani, {\it Topics in Optimal Transportation}, American Mathematical Society, 2003.
\bibitem{WassHEP}
P. Komiske, E. Metodiev, and J. Thaler, ``Metric Space of Collider Events," Phys. Rev. Lett. {\bf 123}, 041801, 2019.
\bibitem{WassHEP2}
M. Erdmann {\it el}, ``Generating and Refining Particle Detector Simulations Using the Wasserstein Distance in Adversarial Networks," Computing and Software for Big Science 2:4 (2018).
\bibitem{WassHEP3}
T. Cai, J. Cheng, and N. Craig, ``Linearized Optimal Transport for Collider Events," Phys. Rev. D {\bf 102}, 116019 (2020).
\bibitem{MMD3}
B. Sriperumbudur {\it et al}, ``On the Empirical Estimation of Integral Probability Metrics", Electronic Journal of Statistics {\bf 6}, 1550-1599 (2012).
\bibitem{MMD4}
B. Sriperumbudur {\it et al}, ``Non-parametric Estimation of Integral Probability Metrics", in Proc. 2010 IEEE International Symposium on Information Theory, 13-18 June 2010.
\bibitem{MMD1}
B. Sriperumbudur {\it et al}, ``Hilbert Space Embeddings and Metrics on Probability Measures", Journal of Machine Learning Research {\bf 11}, 1517-1561 (2010).
\bibitem{MMD1b}
B. Sriperumbudur, K. Fukumizu, and G. Lanckriet, ``Universality, Characteristic Kernels and RKHS Embeddings of Measures," Journal of Machine Learning Research {\bf 12}, 2389-2410 (2011).
\bibitem{MMD2}
B. Sriperumbudur, ``On the Optimal Estimation of Probability Measures in the Weak and Strong Topologies," Bernoulli {\bf 22}, 1839-1893 (2016).
\bibitem{MMD2b}
C. J. Simon-Gabriel and B. Sch{\"o}lkopf, ``Kernel Distribution Embeddings:  Universal Kernels, Characteristic Kernels and Kernel Metrics on Distributions," Journal of Machine Learning Research {\bf 19}, 1-29 (2018).
\bibitem{MMD2c}
C. J. Simon-Gabriel {\it et al}, ``Metrizing Weak Convergence with Maximum Mean Discrepancies" (2021), \url{https://arxiv.org/abs/2006.09268}.
\bibitem{EdistMMD}
D. Sejdinovic {\it et al}, ``Equivalence of Distance-Based and RKHS-Based Statistics in Hypothesis Testing", The Annals of Statistics {\bf 41}, 2263-2291 (2013).
\bibitem{Edist}
G. J. Sz\'ekely and M. L. Rizzo, ``Energy statistics:  A class of statistics based on distances," Journal of Statistical Planning and Inference {\bf 143}, 1249-1272 (2013).
\bibitem{RFF2}
J. Zhao and D. Meng, ``FastMMD:  Ensemble of circular discrepancy for efficient two-sample test", Neural Computation {\bf 27}, 1345-1372 (2015).
\bibitem{RFF3}
A. Rahimi and B. Recht, ``Random Features for Large-Scale Kernel Machines," Adv. Neural Inf. Process. Syst. {\bf 20} (2007).
\bibitem{HSICb}
A. Gretton {\it et al}, ``Measuring Statistical Dependence with Hilbert-Schmidt Norms", Lecture Notes in Computer Science, pp. 63-77 (2005).
\bibitem{HSIC2}
A. Gretton {\it et al}, ``A Kernel Statistical Test of Independence", in Proc. of the 20th International Conference on Neural Information Processing Systems, p. 585-592 (2007).
\bibitem{dCor}
G. Sz\'ekely, G. Rizzo, and M. Bakirov, ``Measuring and testing dependence by correlation of distances", Ann. Stat. {\bf 35}, 2769-2794 (2007).
\bibitem{dCorMetric}
R. Lyons, ``Distance Covariance in Metric Spaces", The Annals of Probability {\bf 41}, 3284-3305 (2013).
\bibitem{RFF}
Q. Zhang {\it et al}, ``Large-scale kernel methods for independence testing", Stat. Comput. {\bf 28}, 113-130 (2018).
\bibitem{JMathPhys}
C. Mitchell, ``Weak Convergence to Equilibrium of Statistical Ensembles in Integrable Hamiltonian Systems", Journal of Mathematical Physics {\bf 60}, 052702 (2019).
\bibitem{Meller}
R. E. Meller {\it et al}, ``Decoherence of Kicked Beams," Technical Note SSC-N-360 (1987), \url{https://lss.fnal.gov/archive/other/ssc/ssc-n-360.pdf}
\bibitem{Walters}
P. Walters, {\it An Introduction to Ergodic Theory}, Springer-Verlag, New York, 1982.
\bibitem{Qiang}
J. Qiang, ``Symplectic multiparticle tracking model for self-consistent space-charge simulation," Phys. Rev. Accel. Beams {\bf 20}, 014203 (2017).
\bibitem{WassSmooth}
S. Nietert, Z. Goldfeld, K. Kato, ``Smooth $p$-Wasserstein distance:  structure, empirical approximation, and statistical applications," Proceedings of the 38th International Conference on Machine Learning, PMRL 139, 8172 (2021).
\bibitem{Wendland}
H. Wendland, ``Piecewise polynomial, positive definite and compactly supported radial functions of minimal degree," Adv. Comput. Math. {\bf 4}, 389-396 (1995).
\bibitem{Wendland2}
A. Chernih and S. Hubbert, ``Closed form representations and properties of the generalised Wendland functions," J. Approx. Theory {\bf 177}, 17-33 (2014).
\bibitem{Gaussian}
H. Minh, ``Some Properties of Gaussian Reproducing Kernel Hilbert Spaces and Their Implications for Function Approximation and Learning Theory", Constructive Approximation {\bf 32}, 307-338 (2010).
\bibitem{MixDef}
This is the definition of {\it strong mixing}.  It is assumed that $A$ and $B$ are ``nice" (Borel) subsets of the phase space $\mathbb{R}^d$, and the map $\mathcal{M}$ is Borel-measurable.  This holds, for example, if $\mathcal{M}$ is continuous.
\end{thebibliography}
\end{document}